\documentclass[sigconf,authorversion,nonacm]{acmart}
\usepackage{balance}
\usepackage[utf8]{inputenc}%(only for the pdftex engine)
\def\BibTeX{{\rm B\kern-.05em{\sc i\kern-.025em b}\kern-.08emT\kern-.1667em\lower.7ex\hbox{E}\kern-.125emX}}

\usepackage{longtable}
\usepackage{tabularx}
\usepackage{afterpage}
\usepackage{placeins}
\usepackage{rotating}
\usepackage{array}
\usepackage{todonotes}
\usepackage{xspace}
\usepackage[linesnumbered,ruled,vlined]{algorithm2e}
\usepackage{adjustbox}
\usepackage[english]{babel}
\usepackage{soul}
\usepackage{flushend}
\usepackage{listings}
\usepackage{amsfonts}
\usepackage{graphicx}
\usepackage{subcaption}
\usepackage{multirow}
\usepackage{latexsym}
\usepackage{fmtcount}
\usepackage{courier}

\theoremstyle{definition}

\usepackage{amsmath}

\DeclareMathOperator*{\argminA}{arg\,min}

\usepackage{hyperref}
\usepackage{url}

\usepackage{breakurl}
\usepackage{cleveref}
\Crefformat{figure}{Fig.~#2#1#3}
\Crefmultiformat{figure}{Figs.~#2#1#3}{ and~#2#1#3}{, #2#1#3}{ and~#2#1#3}
\newcommand{\sysname}{EG-Booster\xspace}

\makeatletter
\patchcmd{\@makecaption}
  {\scshape}
  {}
  {}
  {}

\newcommand*{\da@rightarrow}{\mathchar"0\hexnumber@\symAMSa 4B }
\newcommand*{\da@leftarrow}{\mathchar"0\hexnumber@\symAMSa 4C }
\newcommand*{\xdashrightarrow}[2][]{%
  \mathrel{%
    \mathpalette{\da@xarrow{#1}{#2}{}\da@rightarrow{\,}{}}{}%
  }%
}
\newcommand{\xdashleftarrow}[2][]{%
  \mathrel{%
    \mathpalette{\da@xarrow{#1}{#2}\da@leftarrow{}{}{\,}}{}%
  }%
}
\newcommand*{\da@xarrow}[7]{%
  \sbox0{$\ifx#7\scriptstyle\scriptscriptstyle\else\scriptstyle\fi#5#1#6\m@th$}%
  \sbox2{$\ifx#7\scriptstyle\scriptscriptstyle\else\scriptstyle\fi#5#2#6\m@th$}%
  \sbox4{$#7\dabar@\m@th$}%
  \dimen@=\wd0 %
  \ifdim\wd2 >\dimen@
    \dimen@=\wd2 %   
  \fi
  \count@=2 %
  \def\da@bars{\dabar@\dabar@}%
  \@whiledim\count@\wd4<\dimen@\do{%
    \advance\count@\@ne
    \expandafter\def\expandafter\da@bars\expandafter{%
      \da@bars
      \dabar@ 
    }%
  }%  
  \mathrel{#3}%
  \mathrel{%   
    \mathop{\da@bars}\limits
    \ifx\\#1\\%
    \else
      _{\copy0}%
    \fi
    \ifx\\#2\\%
    \else
      ^{\copy2}%
    \fi
  }%   
  \mathrel{#4}%
}
\makeatother

\newcolumntype{M}[1]{>{\centering\arraybackslash}p{#1}}

\usepackage{enumitem}
\setlist[itemize,1]{leftmargin=1.5\parindent, itemsep=0ex, topsep=0.5ex, % bottomsep=0.5ex
}
\setlist[enumerate,1]{leftmargin=2\parindent, itemsep=0ex, topsep=0.5ex, % bottomsep=0.5ex
}

\setcopyright{none} % switch off copyright text
\begin{document}

\title{EG-Booster: Explanation-Guided Booster of ML Evasion Attacks}

\begin{abstract}
The widespread usage of machine learning (ML) in a myriad of domains has raised questions about its trustworthiness in security-critical environments. Part of the quest for trustworthy ML is robustness evaluation of ML models to test-time adversarial examples. Inline with the trustworthy ML goal, a useful input to potentially aid robustness evaluation is feature-based explanations of model predictions.

In this paper, we present a novel approach called EG-Booster that leverages techniques from explainable ML to guide adversarial example crafting for improved robustness evaluation of ML models before deploying them in security-critical settings. The key insight in EG-Booster is the use of feature-based explanations of model predictions to guide adversarial example crafting by adding consequential perturbations likely to result in model evasion and avoiding non-consequential ones unlikely to contribute to evasion. EG-Booster is agnostic to model architecture, threat model, and supports diverse distance metrics used previously in the literature. 

We evaluate EG-Booster using image classification benchmark datasets, MNIST and CIFAR10. Our findings suggest that EG-Booster significantly improves evasion rate of state-of-the-art attacks while performing less number of perturbations. Through extensive experiments that covers four white-box and three black-box attacks, we demonstrate the effectiveness of EG-Booster against two undefended neural networks trained on MNIST and CIFAR10, and another adversarially-trained ResNet model trained on CIFAR10.
%by $28.78\%$, in average, on a CNN model trained on MNIST and $6.23\%$ increase rate on a CNN model trained on CIFAR10. Moreover, we observe ?.?\% increase in the evasion rate of an adversarially trained ResNet50 model on CIFAR10. In average, across all studied models, we notice a decrease of $\sim 46,39\%$ in the number of feature perturbation performed per sample compared to the baseline attacks. 
Furthermore, we introduce a stability assessment metric and evaluate the reliability of our explanation-based approach by observing the similarity between the model's classification outputs across multiple runs of EG-Booster. Our results across $10$ different runs suggest that EG-Booster's output is stable across distinct runs. Combined with state-of-the-art attacks, we hope EG-Booster will be used as a benchmark in future work that evaluate robustness of ML models to evasion attacks. 
\end{abstract}

\author{Abderrahmen Amich}
\affiliation{%
 \institution{University of Michigan, Dearborn}
 %\city{New York City}
  % \state{NY}
   \country{}
   }
 \email{aamich@umich.edu}

\author{Birhanu Eshete}
\affiliation{%
  \institution{University of Michigan, Dearborn}
  \country{}}
  \email{birhanu@umich.edu}

\settopmatter{printfolios=true}
\maketitle

\section{Introduction}\label{sec: intro}

Machine Learning (ML) models are vulnerable to test-time evasion attacks called {\em adversarial examples} --adversarially-perturbed inputs aimed to mislead a deployed ML model~\cite{FGSM}. Evasion attacks have been the subject of recent research~\cite{FGSM,BIM,PGSM,MIM,CW,HSJA20,uesato2018adversarial} that led to understanding the potential threats posed by these attacks when ML models are deployed in security-critical settings such as self-driving vehicles~\cite{DL-autnonmous17}, malware classification ~\cite{MalConvEvade18}, speech recognition~\cite{DL-Speech2012}, and natural language processing~\cite{NLP-16}.
%For an adversary whose goal is to fool a model, it suffices to obtain $x'$ that is close enough to $x$ according to some distance metric (e.g., one of the $L_p$ norms) and perturbation upper bound $\epsilon$.

Given a ML model $f$ and an input $x$ with a true label $y_{true}$, the goal of a typical evasion attack is to perform minimal perturbations to $x$ and obtain $x'$ similar to $x$ such that $f$ is fooled to misclassify $x'$ as $y' \ne y_{true}$. For a defender whose goal is to conduct pre-deployment robustness assessment of a model to adversarial examples, one needs to adopt a reliable input crafting strategy that can reveal the potential security limitations of the ML system. In this regard, a systematic examination of the suitability of each feature as a potential candidate for adversarial perturbations can be guided by the contribution of each feature in the classification result~\cite{amich2021explanationguided}. In this context, recent progress in feature-based ML explanation techniques~\cite{LIME,SHAP,LEMNA} is interestingly positioned inline with the robustness evaluation goal. In fact, ML explanation methods have been recently utilized to guide robustness evaluation of models against backdoor poisoning of ML models ~\cite{EG-Poisoning}, model extraction~\cite{EG-model-extraction19,EG-model-extraction20}, and membership inference attacks ~\cite{EG-mem-inference2021}, which highlights the utility of ML explanation methods beyond ensuring the transparency of model predictions. 

%The vulnerability of ML models to attacks such as adversarial examples has led to the quest for robust ML that can be confidently deployed in security-critical settings. Robustness to adversarial attacks, however, is just one element of the broader quest for trustworthy ML in which models are expected to also be transparent on their decisions, respect societal and legal norms such as fairness, ethics, and privacy.  
% For instance, in addition to the robustness to adversarial attacks,  efforts towards explainable ML play a major role in the quest of trustworthy ML systems.

In this paper, we present \sysname, an explanation-guided evasion booster that leverages techniques from explainable ML~\cite{LIME,SHAP,LEMNA} to guide adversarial example crafting for improved robustness evaluation of ML models before deploying them in security-critical settings. Inspired by a case study in~\cite{amich2021explanationguided}, our work is the first to leverage black-box model explanations as a guide for systematic robustness evaluation of ML models against adversarial examples. The key insight in \sysname is the use of feature-based explanations of model predictions to guide adversarial example crafting. Given a model $f$, a $d$-dimensional input sample $x$, and a true prediction label $y_{true}$ such that $f(x) = y_{true}$, a ML explanation method returns a weight vector $W_{x,y_{true}} = [w_1, . . ., w_d]$ where each $w_i$ quantifies the contribution of feature $x_i$ to the prediction $y_{true}$. The sign of $w_i$ represents the \textit{direction} of feature $x_i$ with respect to $y_{true}$. If $w_i > 0$, $x_i$ is directed towards $y_{true}$ (in this case we call $x_i$ a {\em positive feature}). In the opposite case, i.e., if $w_i <0$, $x_i$ is directed away from $y_{true}$ (in this case we call $x_i$ a {\em negative feature}). \sysname leverages signs of individual feature weights in two complementary ways. First, it uses the positive weights to identify positive features that are worth perturbing and {\em introduces consequential perturbations} likely to result in $x'$ that will be misclassified as $f(x') = y' \ne y_{true}$. Second, it uses negative weights to identify negative features that need not be perturbed and {\em eliminates non-consequential perturbations} unlikely to contribute to the misclassification goal. \sysname is agnostic to model architecture, adversarial knowledge and capabilities (e.g., black-box, white-box), and supports diverse distance metrics (e.g., common $||.||_p$ norms) used previously in the adversarial examples literature.
% We recall that a ML explainer takes as input a ML model $f$, a sample $x$ and the true prediction $y_{true}$, and attributes for each feature $x_i$, a weight $w_i$ that reflects the importance of $x_i$ in the prediction $f(x)=y_{true}$. . \sysname detects all features that are directed to the correct prediction ($w_i>0$). These features are called \textit{positive features} since they guide the model to make a correct classification. Since the adversary's goal is to cause as much misclassifications as possible, the key intuition is to focus only on perturbing positive features. These features are expected to be \textit{consequential} to the evasion result. 

In an orthogonal line of research that studied explanation {\em stability} and adversarial {\em robustness} of model explanations, some limitations of explanation methods have been documented~\cite{explainEval2020,explainEval}. Recognizing these potential limitations which entail systematic vetting of the reliability of ML explanation methods before using them for robustness assessment of ML models, we introduce an explanation assessment metric called {\em $k$-Stability}, which measures the average stability of evasion results based on the similarities of the target model's predictions returned by multiple runs of \sysname. 

We evaluate \sysname through comprehensive experiments on two benchmark datasets (MNIST and CIFAR10), across white-box and black-box attacks, on state-of-the-art undefended and defended models, covering commonly used $||.||_p$ norms. From white-box attacks we use the Fast Gradient Sign Method (FGS)~\cite{FGSM}, the Basic Iterative Method (BIM)~\cite{BIM}, the Projected Gradient Descent method (PGD)~\cite{PGSM}, and the Carlini and Wagner attack (C\&W)~\cite{CW}. From black-box attacks, we use the Momentum Iterative Method (MIM)~\cite{MIM}, the HopSkipJump Attack (HSJA)~\cite{HSJA20}, and the Simultaneous Perturbation Stochastic Approximation (SPSA) attack~\cite{uesato2018adversarial}.

Across all studied models, we observe a significant increase in the evasion rate of studied baseline attacks after combining them with \sysname. Particularly, results show an average increase of $28.78\%$ of the evasion rate across all attacks performed on undefended MNIST-CNN model. Similar findings are observed on undefended CIFAR10-CNN models, and a defended CIFAR10-ResNet model with average evasion evasion rate increase, respectively, of $6.23\%$ and $5.41\%$. In addition to the reduction of the total number of perturbations compared to the baseline attacks, these findings prove that ML explanation methods can be harnessed to guide ML evasion attacks towards more consequential perturbations. Furthermore, the stability analysis results show that, \sysname's outputs are stable across different runs. Such findings highlight reliability of explanation methods to boost evasion attacks, despite their stability concerns. More detailed results are discussed in section \ref{sec: eval}.

In summary, this paper makes the following contributions:
\begin{itemize}

\item \textbf{Explanation-Guided Evasion Booster.} We introduce the first approach that leverages ML explanation methods towards evaluating robustness of ML models to adversarial examples.

\item \textbf{Stability Analysis.} We introduce a novel stability metric that enables vetting the reliability of ML explanation methods before they are used to guide ML robustness evaluation.

\item \textbf{Comprehensive Evaluation.} We conduct comprehensive evaluations on two benchmark datasets, four white-box attacks, three black-box attacks, different $||.||_p$ distance metrics, on undefended and defended state-of-the-art target models. 

\end{itemize}

To enable reproducibility, we have made available our source code with directions to repeat our experiments. \sysname code is available at:
{\color{blue} \url{ https://github.com/EG-Booster/code}}.

\section{Background}\label{sec: bground}
In this section, we introduce ML explanation methods and ML evasion attacks.

\subsection{ML Explanation Methods}\label{subsec:attribution} 
Humans typically justify their decision by explaining underlying causes used to reach a decision. For instance, in an image classification task (e.g., cats vs. dogs), humans attribute their classification decision (e.g., cat) to certain parts/features (e.g., pointy ears, longer tails) of the image they see, and not all features have the same importance/weight in the decision process. ML models have long been perceived as black-box in their predictions until the advent of explainable ML~\cite{LIME,DeepLIFT,SHAP}, which attribute a decision of a model to features that contributed to the decision. This notion of attribution is based on quantifiable contribution of each feature to a model's decision. Intuitively, an explanation is defined as follows: given an input vector $x = (x_1, . . . , x_d)$, a model $f$, and a prediction label $f(x) = y$, an
explanation method determines why input $x$ has been assigned the label $y$. This explanation is typically represented as a weight vector $W_{x,y} =(w_1, . . . , w_d)$ that captures contribution of each feature $x_i$ towards $f(x) = y$.

ML explanation is usually accomplished by training a substitute model based on the input feature vectors and output predictions of the model, and then use the coefficients of that model to approximate the importance and {\em direction} (class label it leans to) of a feature.  A typical substitute model for explanation is of the form:
% \begin{equation}
    $s(x) = w_{0} +\sum_{i=1}^{d} w_{i}x_{i}$,
% \end{equation}
where $d$ is the number of features, $x$ is the sample, $x_{i}$ is the $i^{th}$ feature for sample $x$, and $w_{i}$ is the weight of feature $x_{i}$ to the model’s decision. While ML explanation methods exist for white-box \cite{Whitebox-exp13,Whitebox-exp14} or black-box~\cite{LIME,LEMNA,SHAP} access to the model, in this work we consider ML explanation methods that have black-box access to the ML model, among which the notable ones are LIME \cite{LIME}, SHAP \cite{SHAP} and LEMNA \cite{LEMNA}. Next, we briefly introduce these explanation methods.

\textbf{LIME and SHAP.}\label{LIME_SHAP}
Ribeiro et al. \cite{LIME} introduce LIME as one of the first model-agnostic black-box methods for locally explaining model output. Lundberg and Lee further extended LIME by proposing SHAP \cite{SHAP}. Both methods approximate the decision function $f_b$ by creating a series of $l$ perturbations of a sample $x$, denoted as $x'_1, . . . , x'_l$ by randomly setting feature values in the vector $x$ to $0$. The methods then proceed by predicting a label $f_b(x'_i)=y_i$ for each $x'_i$ of the $l$ perturbations. This sampling strategy
enables the methods to approximate the local neighborhood of $f_b$ at the point $f_b (x)$. LIME approximates the decision boundary by a weighted linear regression model using Equation \ref{eq:LIME}.
\begin{equation}\label{eq:LIME}
    \argminA_{g \in G} \sum_{i=1}^{l} \pi_x(x'_i)(f_b(x'_i)-g(x'_i))^2
\end{equation}
In Equation \ref{eq:LIME}, $G$ is the set of all linear functions and $\pi_x$ is a function indicating the difference between the input $x$ and a perturbation $x'$. SHAP follows a similar approach but employs the SHAP kernel as weighting function $\pi_x$, which is computed using the \textit{Shapley Values} \cite{shapley} when solving the regression. Shapley Values are a concept from game theory where the features act as players under the objective of finding a fair contribution of the features to the payout --in this case the prediction of the model.

\textbf{LEMNA.}\label{LEMNA}
Another black-box explanation method specifically designed to be a better fit for non-linear models is LEMNA \cite{LEMNA}. As shown in Equation \ref{eq:LEMNA}, it uses a mixture regression model for approximation, that is, a weighted sum of $K$ linear models.
\begin{equation}\label{eq:LEMNA}
    f(x)= \sum_{j=1}^{K} \pi_j(\beta_j . x +\epsilon_j)
\end{equation}
In Equation \ref{eq:LEMNA}, the parameter $K$ specifies the number of models, the random variables $\epsilon = (\epsilon_1, ... , \epsilon_K)$ originate from a normal distribution $\epsilon_i \sim N(0, \sigma)$ and $\pi = (\pi_1, ... , \pi_K)$ holds the weights for each model. The variables $\beta_1, ... , \beta_K$ are the regression coefficients and can be interpreted as $K$ linear approximations of the decision boundary near $f_b(x)$.\\

\textbf{Evaluation of ML explanation methods.} Recent studies have focused on evaluating ML explainers for security applications \cite{explainEval,explainEval2020}. Authors proposed security-related evaluation criteria such as \textit{Accuracy}, \textit{Stability} and \textit{Robustness}. Results have shown that learning-based approach such as LIME, SHAP and LEMNA suffer from unstable output. In particular, their explanations can slightly differ between two distinct runs. In this work, we take measures to ensure that EG-Booster does not inherit the potential stability limitation of the employed ML explanation method (more in Section \ref{stab-ana}). Furthermore, recent studies \cite{exp-Rob1,exp-Rob2} have demonstrated that the explanation results are sensitive to small systematic feature perturbations that preserve the predicted label. Such attacks can potentially alter the explanation results, which raises concerns about the robustness of ML explanation methods. {\em In \sysname, baseline attacks and explanation methods are locally integrated into the robustness evaluation workflow (hence not disclosed to an adversary)}.

\subsection{ML Evasion Attacks}\label{subsec:test-time}
%Two complementary approaches dictate the space of automated malware analysis and detection. {\em Static analysis} techniques examine samples (e.g., Windows PEs) and extract features without executing the samples. On the contrary, {\em dynamic analysis} methods execute binaries to let the behavior unfold, and use behavioral artefacts to classify them. While feature engineering based ML malware classifiers continue to be popular in production malware/intrusion detection systems, recent advances in deep learning have paved the way for raw-byte based ML malware classifiers such as MalConv~\cite{malconv18}. Although ML malware classifiers have been widely employed in practice, akin to models in other domains (e.g., image, text, audio, video), they have been shown to be vulnerable to evasion attacks. In this work, we focus on evasion in a black-box setting for both dynamic and static analysis based ML malware classifiers with feature extraction. In particular, we consider supervised learning for malware classification models, where input samples are Windows PEs and the output is the class label ({\tt benign} or {\tt malware}).

Given a ML model (e.g., image classifier, malware classifier) with a decision function $f:X \rightarrow Y$ that maps an input sample $x \in X$ to a true class label $y_{true} \in Y$, $x'$ = $x  + \delta$ is called an {\em adversarial sample} with an {\em adversarial perturbation} $\delta$ if: 
%\begin{equation}
    $f(x') = y' \ne y_{true}, ||\delta|| < \epsilon$,
%\end{equation}
where $||.||$ is a distance metric (e.g., one of the $L_{p}$ norms)  and $\epsilon$ is the maximum allowable perturbation that results in misclassification while preserving semantic integrity of $x$. Semantic integrity is domain and/or task specific. For instance, in image classification, visual imperceptibility of $x'$ from $x$ is desired while in malware detection $x$ and $x'$ need to satisfy certain functional equivalence (e.g., if $x$ was a malware pre-perturbation, $x'$ is expected to exhibit maliciousness post-perturbation as well). In {
\em untargeted} evasion, the goal is to make the model misclassify a sample to any different class (e.g., for a roadside sign detection model: misclassify red light as any other sign). When the evasion is {\em targeted}, the goal is to make the model to misclassify a sample to a specific target class (e.g., in malware detection: misclassify malware as benign).

Evasion attacks can be done in {\em white-box} or {\em black-box} setting. Most gradient-based evasion techniques ~\cite{FGSM,BIM,PGSM,CW} are white-box because the adversary typically has access to model architecture and parameters/weights, which allows to query the model directly to decide how to increase the model’s loss function. Gradient-based strategies assume that the adversary has access to the gradient function of the model (i.e., white-box access). The core idea is to find the perturbation vector $\delta^\star \in \mathbb{R}^d$ that maximizes the loss function $J(\theta, x, y_{target})$ of the model $f$, where $\theta$ are the parameters (i.e., weights) of the model $f$. In recent years, several white-box adversarial sample crafting methods have been proposed, specially for image classification tasks. Some of the most notable ones are: Fast Gradient Sign Method (FGS)~\cite{FGSM}, Basic Iterative Method (BIM)~\cite{BIM}, Projected Gradient Descent (PGD) method~\cite{PGSM}, and Carlini \& Wagner (C\&W) method~\cite{CW}. Black-box evasion techniques (e.g., MIM~\cite{MIM}, HSJA~\cite{HSJA20}, SPSA~\cite{uesato2018adversarial}) usually start from some initial perturbation $\delta_{0}$, and subsequently probe $f$ on a series of perturbations $f (x + \delta_{i})$, to craft $x'$ such that $f$ misclassifies it to a label different from its original. 

% A wide range of attacks were proposed in this area. Some of the most notable works are: Fast Gradient Sign Method (FGSM)~\cite{FGSM}, Basic Iterative Method (BIM)~\cite{BIM} and Projected Gradient Descent (PGD) method~\cite{PGSM}, and Carlini \& Wagner (C\&W) method~\cite{CW}.  

In Section \ref{sec:attacks}, we briefly introduce the seven reference evasion attacks we used as baselines for \sysname. 
% In the following, we succinctly highlight  some of the well-known evasion methods in the image domain.

\section{Studied Baseline Attacks}\label{sec:attacks}
In this section, we succinctly highlight the four white-box and three black-box attacks we use as baselines for \sysname.

\subsection{White-Box Attacks}
\textbf{Fast-Gradient Sign Method (FGS)} ~\cite{FGSM} is a fast one-step method that crafts an adversarial example. Considering the dot product of the weight vector $\theta$ and an adversarial example (i.e., $x' = x+\delta$), $\theta^\top x' = \theta^\top x + \theta^\top \delta$, the adversarial perturbation causes the activation to grow by $\theta^\top \delta$. Goodfellow et al.~\cite{FGSM} suggested to maximize this increase subject to the maximum perturbation constraint $||\delta||< \epsilon$ by assigning $\delta = sign(\theta)$. Given a sample $x$, the optimal perturbation is given as follows:
\begin{equation}
\label{eq:FGSM}
    \delta^\star = \epsilon . sign(\nabla_xJ(\theta, x, y_{target}))
\end{equation}

\textbf{Basic Iterative Method (BIM)} ~\cite{BIM} was introduced as an improvement of FGS. In BIM, the authors suggest applying the same step as FGS multiple times with a small step size and clip the pixel values of intermediate results after each step to ensure that they are in an $\xi$-neighbourhood of the original image. Formally, the generated adversarial sample after $n+1$ iterations is given as follows:
\begin{equation}
\label{eq:BIM}
\begin{array}{rrclcl}
    x'_{n+1} = Clip_{x,\xi}\{x'_{n} + \epsilon . sign(\nabla_xJ(\theta, x'_n, y_{target})\} \\
    \textrm{s.t.} \quad x'_0 = x\\
\end{array}
\end{equation}

\textbf{Projected Gradient Descent (PGD)}~\cite{PGSM} is basically the same as BIM attack. The only difference is that PGD initializes the example to a random point in the ball of interest \footnote{A ball is the volume space bounded by a sphere; it is also called a solid sphere} (i.e., allowable perturbations decided by the $||.||_\infty$ norm) and does random restarts, while BIM initializes to the original point $x$.

\textbf{Carlini-Wagner (C\&W)}~\cite{CW} is one of the most powerful attacks, where the adversarial example generation problem is formulated as the following optimization problem:
\begin{equation}
\label{eq:BIM}
\begin{array}{rrclcl}
    minimize \quad D(x,x+\delta)\\
    \textrm{s.t.} \quad f(x+\delta) = y_{target}\\
    \quad x+\delta \in [0,1]^n
\end{array}
\end{equation}
The goal is to find a small change $\delta$ such that when added to an image $x$, the image is misclassified (to a targeted class $y_{target}$) by the model but the image is still a valid image. $D$ is some distance metric (e.g $||.||_0$, $||.||_2$ or $||.||_\infty$). Due to the non-linear nature of the classification function $f$, authors defined a simpler objective function $g$ such that $f(x+\delta)=y_{target}$ if and only if $g(x+\delta)<0$. Multiple options of the explicit definition of $g$ are discussed in the paper ~\cite{CW} (e.g., $g(x')=-J(\theta,x',y_{target})+1$). Considering the $p^{th}$ norm as the distance $D$, the optimization problem is simplified as follows:
\begin{equation}
\label{eq:BIM}
\begin{array}{rrclcl}
    minimize \quad ||\delta||_p + c.g(x+\delta)\\
    \textrm{s.t.} \quad x+\delta \in [0,1]^n\\
\end{array}
\end{equation}
where c > 0 is a suitably chosen constant.\\

\subsection{Black-Box Attacks}

\textbf{Momentum Iterative Method (MIM)~\cite{MIM}} is proposed as a technique for addressing the likely limitation of BIM: the greedy move of the adversarial example in the direction of the gradient in each iteration can easily drop it into poor local maxima. It does so by leveraging the momentum method ~\cite{momentum} that stabilizes gradient updates by accumulating a velocity vector in the gradient direction of the loss function across iterations, and allows the algorithm to escape from poor local maxima. For a decay factor $\mu$ the gradient $\mathbf{g}$ for iteration $t+1$ is updated as follows:
\begin{equation}
\label{eq:MIM}
\mathbf{g}_{t+1} = \mu \mathbf{g}_{t}+ \frac{\nabla_xJ_{\theta}(x'_t, y)}{||\nabla_xJ_{\theta}(x'_t, y)||_1},
\end{equation}
where the adversarial example is obtained as  $x'_{t+1} = x'_t +  \epsilon . sign(g_{t+1})$.

\textbf{HopSkipJump Attack (HSJA)~\cite{HSJA20}} is an attack that relies on prediction label only to create an adversarial example. The key intuition of HSJA is based on gradient-direction estimation, motivated by zeroth-order optimization. Given an input $x$, first the gradient direction estimation is computed as follows:
\[
S_{x}(x')=
\begin{cases}
    \max_{y\ne y_{true}} f_y(x') - f_{y_{true}}(x') & \text{(Untargeted})\\
    f_{y_{target}}(x') -  \max_{y\ne_{y_{target}} f_y(x')}           & \text{(Targeted)}
\end{cases}
\]

A perturbed input $x'$ is a successful attack if and only if $S_{x}(x') > 0$. The boundary between successful and unsuccessful perturbed inputs is given by bd($S_x$)  = $ z \in [0,1]^d \quad|\quad S_x(z)=0$. 
As an indicator of successful perturbation, the authors introduce the Boolean-valued function $\phi_x : [0, 1]^d \xrightarrow{} \{-1, 1\}$ via:
\[
\phi_{x}(x')= \text{sign}(S_{x}(x')) = 
\begin{cases}
    1& \textrm{if} \quad S_{x}(x') >0\\
    -1 & \text{Otherwise.}
\end{cases}
\]
The goal of an adversarial attack is to generate a perturbed sample $x'$
such that $\phi_{x}(x') = 1$, while keeping $x'$ close to the original input sample $x$. This can be formulated as an optimization problem: $\min_{x'} d(x',x)$ such that $\phi_{x}(x') =1$, where $d$ is a distance metric that quantifies similarity.\\

\noindent \textbf{Simultaneous Perturbation Stochastic Approximation (SPSA)} by Uesato et al. ~\cite{uesato2018adversarial} is an attack that re-purposes gradient-free optimization techniques into adversarial example attacks. They explore adversarial risk as a measure of the model’s performance on worst-case inputs. Since the exact adversarial risk is computationally intractable to exactly evaluate, they rather frame commonly used attacks (such as the ones we described earlier) and adversarial evaluation metrics to define a tractable surrogate objective to the true adversarial risk. The details of the algorithm are in~\cite{uesato2018adversarial}.
\section{Explanation-Guided Booster}\label{sec: approach}
In this section, we describe the details of \sysname. 
\begin{algorithm}
%\scriptsize
\KwResult{$x'_{boost}$, $N_{nc}$, $N_{c}$}
\SetAlgoLined
\textbf{Input:}\\
$f$ : black-box classification function;\\
$x$: legitimate input sample;\\
$x'_{baseline}$: $x$'s adversarial variant returned by baseline attack;\\
$W_{x,y_{true}}$: feature weights (explanations) of a $x$ toward $y_{true}$;\\
$||.||_p$: norm used in the baseline attack;\\
$\epsilon$: perturbation bound used in the baseline attack;\\
$max\_iter$: maximum iterations to make bounded perturbation $\delta_i$;\\
\textbf{Initialization:}\\
$W_{x,y_{true}} \gets ML\_Explainer(f, x, y_{true})$;\\
$d \gets dim(x)$;\\
$x'_{boost} \gets x'_{baseline}$;\\
$y_{baseline} \gets f(x'_{baseline}$);\\
$N_{nc}$,$N_{c} \gets 0$;\\
\textbf{Output:}\\
\tcp{Eliminate non-consequential perturbations}
\ForEach {$w_i \in (W_{x,y_{true}},), i \leq d$}
    {
    \If{$w_i<0$}
        {
           $x'_{boost}[i] \gets x[i]$\;
           \eIf{$y_{baseline} \ne y_{true} \and f(x'_{boost})$ = $y_{true}$}
                {$x'_{boost}[i] \gets x'_{baseline}[i]$}
                {$N_{nc}=N_{nc}+1$}
        }
    }
\tcp{Add consequential perturbations if baseline attack fails}
\ForEach {$w_i \in (W_{x,y_{true}},) , i \leq d$}{
    \If{$w_i>0 \and f(x'_{boost}) = y_{true}$}{
        $\delta_i = get\_delta(x)$;\break
        $x'_{boost}[i] \gets x'_{boost}[i] + \delta_i$\;\break
        $iter \gets 0$;\break
        \While{$||x'_{boost} - x||_p>\epsilon$}{
            $iter \gets iter+1$;\break
            $reduce\_and\_repeat(\delta_i)$;\break
            \If{$iter=max\_iter$}{
            break;}
        }
   \eIf{$||x'_{boost} - x||_p>\epsilon$}{
   $x'_{boost}[i] \gets x'_{baseline}[i]$;
   }
   {$N_{c}=N_{c}+1$
   }
  }
}
\caption{Explanation-Guided Booster Attack.}
\label{alg:EG-Boost}
\end{algorithm}
\subsection{Overview}
In the reference attacks we introduced in Section \ref{sec:attacks}, without loss of generality, the problem of crafting an adversarial example can be stated as follows: Given an input $x \in \mathbb{R}^d$ such that $f(x) = y_{true}$, the goal of the attack is to find the optimal perturbation $\delta^\star$ such that the adversarial example $x' = x + \delta^\star$ is misclassified as $f(x') = y \ne y_{true}$. This problem is formulated as:  
% Evasion attacks are typically defined as a constrained optimization problem (explained in section \ref{subsec:test-time}). For instance, given a deployed ML model (e.g., image classifier) with  a decision function $f:X \rightarrow Y$ that maps an input sample $x \in X$ to a true class label $y_{true} \in Y$, then $x'$ = $x  + \delta$ is called an {\em adversarial sample} with an {\em adversarial perturbation} $\delta$ if: 
% %\begin{equation}
%     $f(x') = y' \ne y_{true}, ||\delta|| < \epsilon$,
% %\end{equation}
% where $||.||$ is a distance metric (e.g., one of the $L_{p}$ norms)  and $\epsilon$ is the maximum allowable perturbation that results in misclassification while preserving semantic integrity of $x$. Formally, the evasion problem is typically defined as follows\\
\begin{equation}
\label{eq:unguided}
\begin{array}{rrclcl}
\displaystyle \delta^\star = \argminA{\delta \in \mathbb{R}^d} & f(x + \delta)\\
\textrm{s.t.} & ||\delta^\star||<\epsilon\\
\end{array}
\end{equation}
The only constraint of Equation \ref{eq:unguided} is that the perturbation size of the vector $\delta$ is bounded by a maximum allowable perturbation size $\epsilon$ (i.e., $||\delta||_p<\epsilon$), which ensures that adversarial manipulations preserve the semantics of the original input. However, it does not guarantee that all single feature perturbations (i.e., $x'_i = x_i + \delta_i$) are \textit{consequential} to result in evasion. Our Explanation-Guided Booster (\sysname) approach improves Equation \ref{eq:unguided} to guide any $L_p$ norm attack to perform only necessary perturbations that can cause evasion. In addition to the upper bound constraint $\epsilon$, \sysname satisfies an additional constraint that guarantees the perturbation of only the features that are initially contributing to a correct prediction (\textit{positive features}). In other words, \sysname serves to guide the state-of-the-art attacks towards perturbing only features with positive explanation weights $w_i>0$ where $w_i \in W_{x,y_{true}} =\{w_1,...,w_d\}$ are the feature weights (explanations) of the input $x$ towards the true label $y_{true}$ (as explained in Section \ref{subsec:attribution}). Formally, the \sysname adversarial example crafting problem is stated as follows:
\begin{equation}
\label{eq:Booster}
\begin{array}{rrclcl}
\displaystyle \delta^\star = \argminA{\delta \in \mathbb{R}^d} & f(x + \delta)\\
\textrm{s.t.} & ||\delta^\star||<\epsilon\\
& W_{x,y_{true}}(\delta^\star) > 0\\
\end{array}
\end{equation}
The second constraint $W_{x,y_{true}}(\delta^\star) >0$ ensures that only features with positive explanation weights are selected for perturbation.

% Being mindful of the \textit{accuracy} and \textit{stability} limitations of ML explanation methods that we discussed in section \ref{subsec:attribution} ~\cite{explainEval,explainEval2020},
As shown in Algorithm \ref{alg:EG-Boost} (line 10), first, \sysname performs ML explanation on the input $x$, in order to gather the original \textit{direction} of each feature $x_i$ with respect to the true label $y_{true}$. Next, using explanation results ($W_{x,y_{true}}$), \sysname reviews initial perturbations performed by a baseline attack by eliminating features that have negative explanation weight (lines 16--25) and adding more perturbations on features that have positive explanation weight (lines 26--44). We ensure that \sysname's intervention is pre-conditioned on maintaining at least the same evasion success of an evasion attack, if not improving it. Particularly, in case the initial adversarial sample generated by a baseline attack $x'_{baseline}$ succeeds to evade the model, \sysname eliminates only perturbations that do not affect the initial evasion result. In this case, even if there exist unperturbed positive features when $x'_{baseline}$ was crafted, \sysname does not perform any additional perturbations since the evasion result has been already achieved.  

%Definitions \ref{def:cons} and \ref{def:non-cons} are used to decide whether a feature is a candidate for perturbation. 
%\begin{definition}[\textbf{Consequential Perturbation}]
%A feature perturbation $x'_i = x_i + \delta_i$ is \textit{consequential} if and only if $x_i$ is directed to $y_{true}$ (i.e., $w_i>0$).
%\label{def:cons}
%\end{definition}
%The intuition behind Definition \ref{def:cons} is that perturbing features that are directed to $y_{true}$ would reduce the number of relevant features that cause a correct classification. Therefore, it is consequential to an untargeted misclassification.
%\begin{definition}[\textbf{Non-Consequential Perturbation}]
%A feature perturbation $x'_i = x_i + \delta_i$ is \textit{non-consequential} if and only if $x_i$ is not directed to $y_{true}$ (i.e., $w_i<0$).
%\label{def:non-cons}
%\end{definition}
%Similar to Definition \ref{def:cons}, the intuition behind Definition \ref{def:non-cons} is that features that are already not directed to the correct prediction are not strong candidates for perturbation. Perturbing such features would not be necessarily consequential to result in misclassification.

Next, we use Algorithm \ref{alg:EG-Boost} to further explain the details of \sysname.
\subsection{Eliminating Non-Consequential Perturbations}
\label{eliminate}
\begin{figure}[t!]
    
    \centering
    % \scalebox{.5}{
    %\framebox{
    \includegraphics[width=\columnwidth]{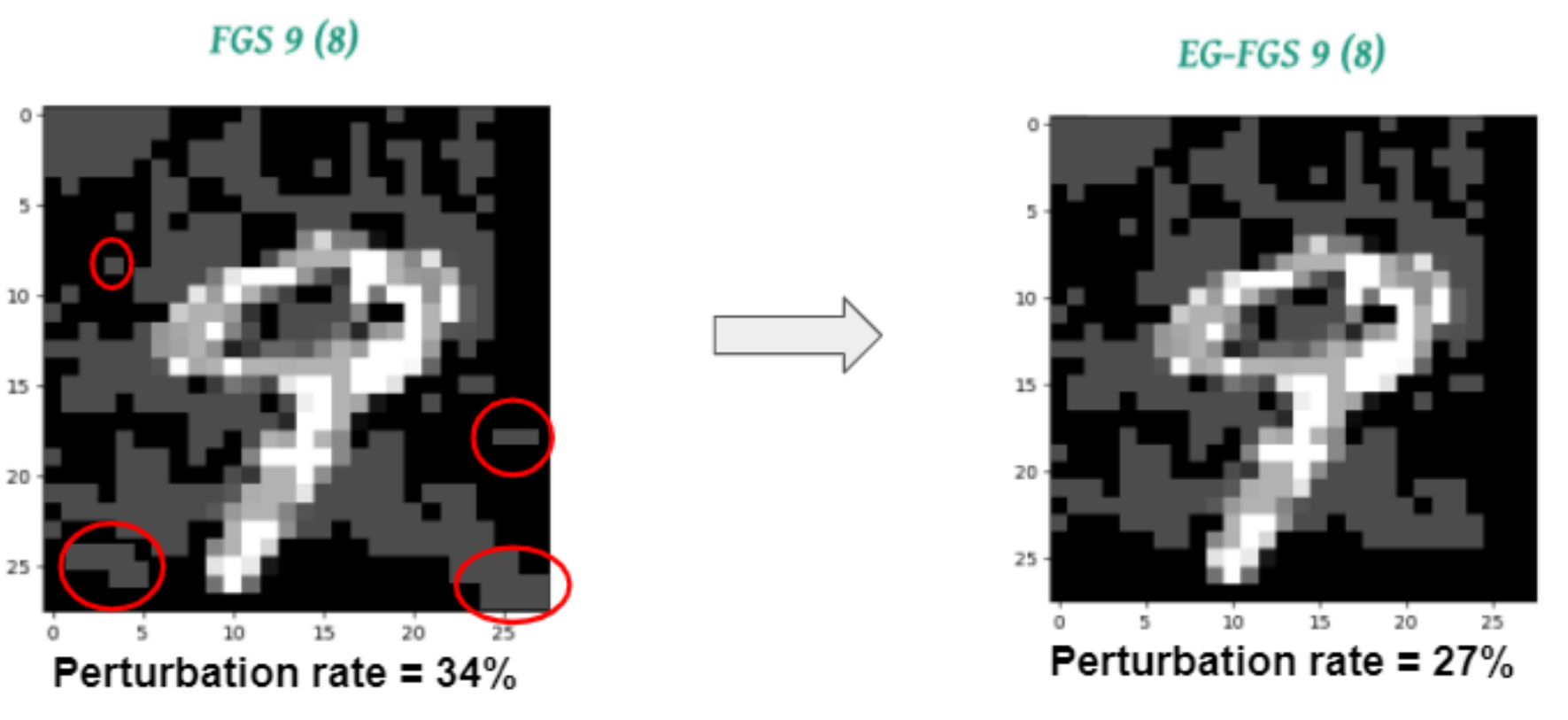}
   % }
   \vspace*{1em}
    \caption{An example of the impact of eliminating non-consequential perturbations from a perturbed MNIST input that already succeeds to evade a CNN model using FGS. We put the new prediction of each version of the input image in parentheses (.).}
    \label{fig:impact_neg}

\end{figure}
As shown in Algorithm \ref{alg:EG-Boost} (lines 16--25), \sysname starts by eliminating unnecessary perturbations that are expected to be \textit{non-consequential} to the evasion result (if any). If the adversarial input $x'_{baseline}$ produced by the baseline attack is already leading to a successful evasion, \sysname ensures that the perturbation elimination step does not affect the initial evasion success (lines 19--20). Accordingly, \sysname intervenes to eliminate the perturbations that have no effect on the final evasive prediction. More precisely, it searches for perturbed features that are originally not directed to the true label $y_{true}$ (line 17), and then it restores the original values of those features (i.e., eliminates the perturbation) (line 18) while ensuring the validity of the initial evasion if it exists (it does so by skipping any elimination of feature perturbations that do not preserve the model evasion). As we will show in our experiments (Section \ref{subsec:perturbation-change}), this step can lead to a significant drop in the total number of perturbed features which results in a smaller perturbation size $||\delta^\star||_p$. 

Figure \ref{fig:impact_neg} shows an MNIST-CNN example of the impact of enhancing the FGS \cite{FGSM} attack with \sysname when the baseline perturbations are already evasive. In this example, FGS perturbations fool the CNN model to incorrectly predict digit `9' as digit `8'. The red circles on the left hand side image show the detection of non-consequential perturbations  signaled by \sysname using the pre-perturbation model explanation results. The new version of the input image shown on the right hand side of the figure illustrates the impact of eliminating those non-consequential feature perturbations. The number of feature perturbations is reduced by 7\% while preserving the evasion success.

The other alternative is when the baseline attack initially fails to evade the model. In such a case, it is crucial to perform this perturbation elimination step over all non-consequential features before making any additional perturbations. Doing so reduces the perturbation size $||\delta||$ which provides a larger gap ($\epsilon - ||\delta||$) for performing additional consequential perturbations that can cause a misclassification.

\subsection{Adding Consequential Perturbations}
\label{add}
\begin{figure}[t!]
    
    \centering
    %\framebox{
    % \scalebox{.5}{
    \includegraphics[width=\columnwidth]{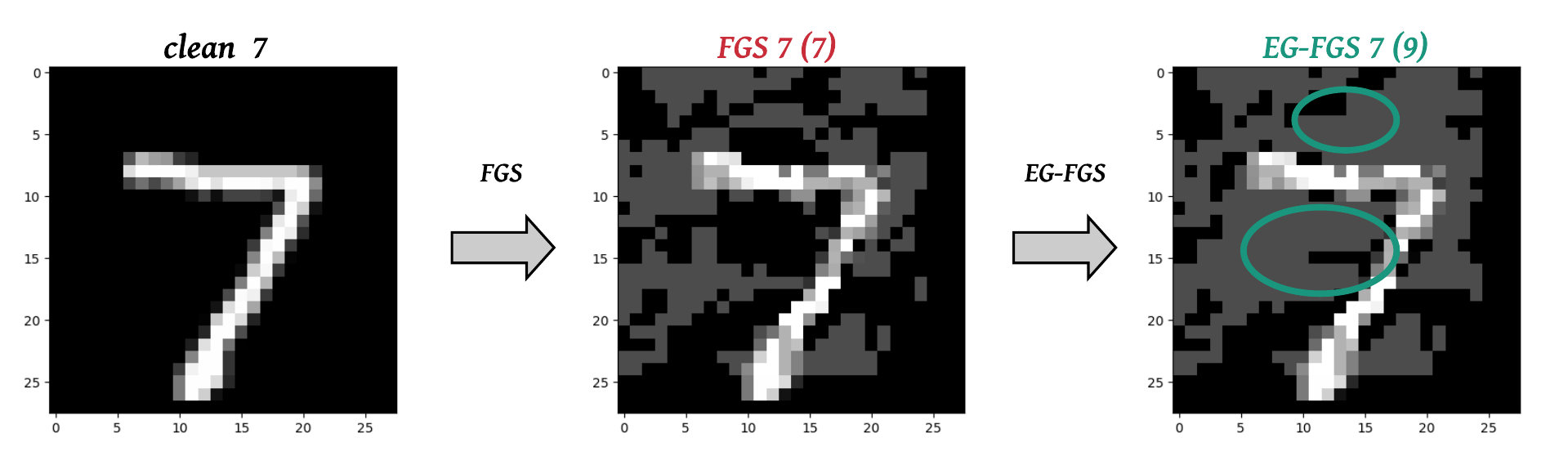}
   % }
   \vspace*{1em}
    \caption{An example of the impact of adding consequential perturbations to a perturbed MNIST input that failed to evade a CNN model using FGS. We put the new prediction of each version of the input image in parentheses (.).}
    \label{fig:impact_pos}
    % \vspace*{-1em}

\end{figure}
This second step is only needed in case the adversarial sample has still failed to evade the model after eliminating non-consequential perturbations. In this case, \sysname starts searching for unperturbed features that are directed towards the true label $y_{true}$. When it finds such features, it then incrementally adds small perturbations $\delta_i$ to these features since they are signaled by the ML explainer as crucial features directed to a correct classification (lines 28--29). The function \textit{get\_delta(x)} carefully chooses a random feature perturbation $\delta_i$ that satisfies the feature constraint(s) of the dataset at hand (line 28). For instance, in image classification, if the feature representation of the samples is in gray-scale, pixel perturbation should be in the allowable range of feature values (e.g., [$0,1$] for normalized MNIST features). In case a feature perturbation $x_i+\delta_i$ breaks the upper bound constraint $||\delta||_p>\epsilon$, \sysname iteratively keeps reducing the perturbation $\delta_i$ until the bound constraint is reached or the number of iterations hits its upper bound (lines 30--37).

In case all features directed to the correct label are already perturbed, \sysname proceeds by attempting to increase the perturbations initially performed by the baseline attack on positive features. It does so while respecting the upper bound constraint ($||\delta||<\epsilon$). This process immediately terminates whenever the evasion goal is achieved.

Figure \ref{fig:impact_pos} shows an MNIST-CNN example of the impact of enhancing the FGS attack \cite{FGSM} with \sysname. In this example, FGS perturbations originally failed to fool the CNN model to predict an incorrect label (i.e., $y \ne$ `7'). The right hand side image shows the perturbations (circled in green) added by \sysname that are consequential to fool the CNN model into predicting the input image as `9' instead of `7'.

\subsection{Stability Analysis}
\label{stab-ana}
As stated earlier (Section \ref{subsec:attribution}), when evaluated on security-sensitive learning tasks (e.g., malware classifiers), ML explanation methods have been found to exhibit output \textit{instability} where by explanation weights of the same input sample $x$ can differ from one run to another~\cite{explainEval,explainEval2020}. In order to ensure that \sysname does not inherit the instability limitation of the ML explainer, we perform stability analysis where we compare the \textit{similarity} of the prediction results returned by the target model $f$ after performing an explanation-guided attack over multiple runs. More precisely, we define the \textit{k-stability} metric that quantifies the stability of \sysname across \textit{k} distinct runs. It measures the average run-wise (e.g., each pair of runs) \textit{similarity} between the returned predictions of the same adversarial sample after \sysname attack. The \textit{similarity} between two runs $i$ and $j$, $similarity(i,j)$, is the intersection size of the predictions returned by the two runs over all samples in the test set (i.e., the number of matching predictions). Formally, we define the \textit{k-stability} as follows.
\begin{equation}
\begin{array}{rrclcl}
\displaystyle 
\textit{k-stability} = \frac{1}{k(k-1)}\sum_{i=1}^{k} \sum_{\substack{j=1 \\ j\ne i}}^{k} similarity(i,j)
\end{array}
\label{eq:k-stab}
\end{equation}
The instability of explanation methods is mainly observed in the minor differences between the magnitude of the returned explanation weights of each feature $||w_i||$ which might sometimes lead to different feature ranking. However, the direction of a feature $x_i$ is less likely to change from one run to another as it is uniquely decided by the sign of its explanation weight $sign(w_i)$. This is mainly visible on images as it is possible to plot feature directions with different colors on top of the original image. \sysname only relies on the sign of the explanation weights $w_i \in W_{x,y_{true}}$ in the process of detecting \textit{non-consequential perturbations} and candidates for \textit{consequential perturbations}. As a result, the output of \sysname is expected to be more stable than output of the underlying explanation method. 

To validate our intuition we compare the stability of \sysname with the stability of the employed explanation method. Thus, we compute the \textit{(k,l)-Stability} metric proposed by a prior work \cite{explainEval2020} that evaluates the stability of ML explanation methods for malware classification models. The \textit{(k,l)-Stability} measure computes the average explanation similarity based on
intersection of features that are ranked in the top-l, returned by separate $k$ runs of an explanation method. More precisely, given two explanation results, $W_{x,y_{true}}(l,i)$ and $W_{x,y_{true}}(l,j)$ returned by two different runs $i$ and $j$, and a parameter $l$, their similarity is obtained based on the Dice coefficient:.
\begin{equation}
\begin{array}{rrclcl}
\displaystyle 
\textit{sim($W_{x,y_{true}}(l,i)$,$W_{x,y_{true}}(l,j)$)} = 2 * \frac{W_{x,y_{true}}(l,i) \cap W_{x,y_{true}}(l,j)}{|W_{x,y_{true}}(l,i)|+|W_{x,y_{true}}(l,j)|}, 
\end{array}
\label{eq:exp-sim}
\end{equation}
where $W_{x,y_{true}}(l,i)$ denotes the top-$l$ features of sample $x$ with respect to the prediction $y_{true}$ returned by run $i$. Over a total of $k$-runs, the average \textit{(k,l)-Stability} on a sample $x$ is defined as follows:
\begin{equation}
\begin{array}{rrclcl}
\displaystyle 
\textit{(k,l)-stability} = \frac{1}{k(k-1)}\sum_{i=1}^{k} \sum_{\substack{j=1 \\ j\ne i}}^{k} sim(W_{x,y_{true}}(l,i),W_{x,y_{true}}(l,j))
\end{array}
\label{eq:kl-stab}
\end{equation}

More empirical discussions about the stability results over all samples in a test set $X_t$ are presented in Section \ref{subsec:stability-results}.
\section{Experimental Evaluation}\label{sec: eval}
In this section we report our findings through systematic evaluation of \sysname. We first describe our experimental setup in Section\ref{subsec:setup}. In Section \ref{subsec:evasion-results}, we present details of \sysname evasion accuracy effectivness. In Sections \ref{subsec:perturbation-change}, \ref{subsec:stability-results}, and \ref{subsec:exec-time}, we present our findings on perturbation change, stability analysis, and execution time of \sysname, respectively.

\subsection{Experimental Setup}\label{subsec:setup}

\textbf{Datasets.}
We evaluate \sysname on two standard datasets used for adversarial robustness evaluation of deep neural networks: MNIST~\cite{MNIST}, a handwritten digit recognition task with ten class labels ($0-9$) and CIFAR10\cite{cifar}, an image classification task with ten classes. We use the $10$K test samples of both datasets to evaluate the performance of \sysname on undefended neural networks (described next). To evaluate \sysname against a defended model, we use $5$K samples of the CIFAR10 test set.

\textbf{Models.}
For MNIST, we train a state of the art 7-layer CNN model from ``Model Zoo''\footnote{https://gitlab.com/secml/secml/-/tree/master/src/secml/model\_zoo}, provided by the SecML library \cite{melis2019secml}. It is composed of 3-conv layers+ReLU, 1-Flatten layer, 1-fully-connected layer+ReLU, a dropout-layer (p=0.5), and finally a Flatten-layer. We call this model \textbf{MNIST-CNN}. It reaches a test accuracy of $98.32\%$ over all $10$K test samples.

For CIFAR10, we consider two different models: a CNN model \cite{CIFAR-CNN} with 4-conv2D and a benchmark adversarially-trained ResNet50 model. In our experimental results, we call the first undefended model \textbf{CIFAR10-CNN} and the second defended model \textbf{CIFAR10-ResNet}. More precisely, the structure of CIFAR10-CNN is: 2-conv2D+ ReLU, 1-MaxPool2D, 1-dropout(p=0.25), 2-conv2D+ReLU, 1-MaxPool2D, 1-dropout(p=0.25), 2-fully-connected+ReLU, 1-dropout(p=0.25), and 1-fully-connected. To avoid the reduction of height and width of the images, padding is used in every convolutional layer. This model reaches a test accuracy of $86.41\%$ after $50$ epochs. It outperforms the benchmark model adopted by Carlini \& Wagner (2017)\cite{CW} and Papernot et al. (2016) \cite{distillation}. 

CIFAR10-ResNet is a state-of-the-art adversarially-trained network model from the"robustness" library \cite{robustness} proposed by MadryLab \footnote{https://github.com/MadryLab/robustness/blob/master/robustness/cifar\_models/ResNet.py}. It is trained on images generated with PGD attack using $L_2$ norm and $\epsilon=0.5$ as perturbation bound. It reaches a test accuracy of $92.36\%$ over all $10$K test set of CIFAR10.

\textbf{Baseline Attacks.}
We evaluate \sysname in a \textit{white-box} setting using 4 state-of-the-art white-box baseline attacks i.e., FGS~\cite{FGSM}, BIM~\cite{BIM}, PGD~\cite{PGSM}, and C\&W~\cite{CW}. Furthermore, we validate \sysname in a \textit{black-box} setting using 3 state-of-the-art black-box attacks: MIM~\cite{MIM}, HSJA~\cite{HSJA20}, and SPSA~\cite{uesato2018adversarial}. Detailed explanations of all the 7 baseline attacks are provided in Section \ref{sec:attacks}. We provide the hyper-parameter choices of each attack in the Appendix (Table \ref{tab:params} in Section \ref{hyper}). 

%\textbf{Attack Hyper-parameters.}

\textbf{ML Explainer.}
The performance of \sysname is influenced by the effectiveness of the employed ML explanation method in detecting the \textit{direction} of each feature $x_i$ in a sample $x$. It is, therefore, crucial to suitably choose the ML explainer according to the studied systems and the deployed models. In our experiments, we focus on image datasets and Neural Network models, therefore, we pick SHAP~\cite{SHAP}, as it is proved to be effective in explaining deep neural networks, especially in the image classification domain. Furthermore, SHAP authors proposed a ML explainer called ``Deep Explainer'', designed for deep learning models, specifically for image classification. SHAP has no access to the target model, which makes \sysname suitable for either black-box or white-box threat model. We note that independent recent studies \cite{explainEval,explainEval2020} evaluated ML explanation methods for malware classifiers. They revealed that LIME outperforms other approaches in security systems in terms of accuracy and stability. Thus, we recommend using LIME for future deployment of \sysname for robustness evaluation of ML malware detectors.

\textbf{Evaluation metrics.}
We use the following four metrics to evaluate \sysname:

\textbf{{\em Evasion Rate}}: First, we quantify the effectiveness of \sysname by monitoring changes to the \textit{percentage of successful evasions} with respect to the total test set, from the baseline attacks to \sysname attacks.

\textbf{{\em Average Perturbation Change}}: Second, we keep track of the average changes to the number of perturbed features using baseline attack versus using \sysname attack to show the impact of the addition or elimination of perturbations performed by \sysname. Referring to Algorithm \ref{alg:EG-Boost} (lines 23 and 42), \sysname keeps track of the number of added perturbations ($N_c$) and the number of eliminated perturbations ($N_{nc}$) per-image. Thus, the total of perturbation changes per-image is $N_{c} - N_{nc}$, and the average perturbation change across all images of the test set ($X_t$) is:
\begin{equation}
\begin{array}{rrclcl}
\displaystyle 
\textit{Average Perturbation Change} = \frac{1}{|X_t|}\sum_{x\in X_t}^{} \frac{N_c(x) - N_{nc}(x)}{N_{baseline}(x)},
\end{array}
\label{eq:pert}
\end{equation}
where $N_{baseline}(x)$ is the total number of perturbations initially performed by the baseline attack on an example $x$.  When \textit{Average Perturbation Change}$>0$, on average, \sysname is performing more perturbation additions than eliminations ($N_c>N_{nc}$). Otherwise, it is performing more elimination of non-consequential perturbations ($N_c<N_{nc}$). The average becomes zero when there are equal number of added and eliminated perturbations.
% Tables \ref{tab:MNIST-CNN}, \ref{tab:CIFAR-CNN}, and \ref{tab:CIFAR-ResNet} summarize the results of these metrics for different models and baseline attacks.

\textbf{{\em k-Stability \& (k,l)-Stability}}: To evaluate the reliability of \sysname, we use the $k-Stability$ (Equation \ref{eq:k-stab}) and $(k,l)-Stability$ (Equation \ref{eq:kl-stab}) measures we introduced in Section \ref{stab-ana}. We compute the $k-Stability$ of \sysname for different $k$ values and we compare it to the value of the $(k,l)-Stability$ of the employed explanation method (i.e., SHAP), using the top-10 features $(l=10)$ and different values of $k$. Both metrics are calculated in average over $1000$ samples.

\textbf{{\em Execution Time}}: Across both datasets and all models, we measure the per-image execution time (in seconds) taken by \sysname for different baseline attacks.
% In Figure \ref{fig:time}, we provide \textit{the execution time} of \sysname per image using the same experiments performed in tables \ref{tab:MNIST-CNN}, \ref{tab:CIFAR-CNN} and \ref{tab:CIFAR-ResNet} across all studied models.

\begin{table*}[t]
\centering

\begin{adjustbox}{angle=0}
\scalebox{.95}{
\begin{tabular}{!{\color{black}\vrule}l!{\color{black}\vrule}l!{\color{black}\vrule}l!{\color{black}\vrule}l!{\color{black}\vrule}l!{\color{black}\vrule}l!{\color{black}\vrule}l!{\color{black}\vrule}l!{\color{black}\vrule}l!{\color{black}\vrule}l!{\color{black}\vrule}l!{\color{black}\vrule}l!{\color{black}\vrule}l!{\color{black}\vrule}!{\color{black}\vrule}l!{\color{black}\vrule}l!{\color{black}\vrule}l!{\color{black}\vrule}l!{\color{black}\vrule}l!{\color{black}\vrule}l!{\color{black}\vrule}l!{\color{black}\vrule}l!{\color{black}\vrule}l!{\color{black}\vrule}l!{\color{black}\vrule}l!{\color{black}\vrule}l!{\color{black}\vrule}l!{\color{black}\vrule}!{\color{black}\vrule}l!{\color{black}\vrule}l!{\color{black}\vrule}l!{\color{black}\vrule}l!{\color{black}\vrule}l!{\color{black}\vrule}l!{\color{black}\vrule}l!{\color{black}\vrule}l!{\color{black}\vrule}l!{\color{black}\vrule}l!{\color{black}\vrule}l!{\color{black}\vrule}l!{\color{black}\vrule}l!{\color{black}\vrule}!{\color{black}\vrule}l!{\color{black}\vrule}l!{\color{black}\vrule}l!{\color{black}\vrule}l!{\color{black}\vrule}l!{\color{black}\vrule}l!{\color{black}\vrule}l!{\color{black}\vrule}l!{\color{black}\vrule}l!{\color{black}\vrule}l!{\color{black}\vrule}l!{\color{black}\vrule}l!{\color{black}\vrule}l!{\color{black}\vrule}} 
\hline
  & \multicolumn{6}{c!{\color{black}\vrule}}{\textbf{Undefended MNIST-CNN}} 
  %&\multicolumn{9}{l!{\color{black}\vrule}}{\textbf{CIFAR10-CNN}}
\\
\hline
\multirow{2}{*}{{\bf Baseline Attacks}}  & 
\multicolumn{5}{c!{\color{black}\vrule}}{White-Box} &
\multicolumn{1}{c!{\color{black}\vrule}}{Black-Box} 
%& \multicolumn{6}{l!{\color{black}\vrule}}{White-Box} &
%\multicolumn{3}{l!{\color{black}\vrule}}{Black-Box}

\\ 
%\hline
\cline{2-7}&  

\multicolumn{1}{l!{\color{black}\vrule}}{\em FGS ($X_t=10K$)}   & \multicolumn{1}{l!{\color{black}\vrule}}{\em BIM ($X_t=10K$)}   & \multicolumn{2}{l!{\color{black}\vrule}}{\em PGD ($X_t=10K$)}   & \multicolumn{1}{l!{\color{black}\vrule}}{\em C\&W ($X_t=5K$)} & \multicolumn{1}{l!{\color{black}\vrule}}{\em MIM ($X_t=10K$)}  \\

\cline{1-7}
{\bf $||.||_p$ Norms}   & {\em $L_\infty$}& {\em $L_\infty$}    & {\em $L_2$} & {\em $L_\infty$}    & {\em $L_2$} & {\em $L_\infty$} \\
\hline\hline
\textbf{Initial Evasion Rate}    

& 28.20\%
& 49.92\% 
& 99.80 \%  
& 55.09\%         
& 99.14\%  
& 41.34\%
\\
\hline
%\cline{1-9}
 \textbf{\sysname Evasion Rate}      
   
 & \textbf{89.71\%}         
 
 & \textbf{89.08\%}
   
 & \textbf{99.89\%}   
 & \textbf{88.69\%}          
 & \textbf{99.72\%} 
 
 & \textbf{79.08\%}
   \\
\hline
%\cline{1-9}
%\cline{1-9}
\textbf{Average Perturbation Change}        
   
& +14.44\%         

& +14.47\%

& \textbf{-5.46\%}   
& \textbf{-4.39\%}          
& \textbf{-15.36\%} 

& \textbf{-5.49\%}
 \\
\hline
%\cline{1-9}
\end{tabular}}
\end{adjustbox}

 \vspace*{2em}
\caption{Summary of \sysname results on undefended MNIST-CNN Model ($\epsilon = 0.3$).}
\label{tab:MNIST-CNN}
\end{table*}

\begin{table*}[t]

\centering

\begin{adjustbox}{angle=0}
\scalebox{.99}{
\begin{tabular}{!{\color{black}\vrule}l!{\color{black}\vrule}l!{\color{black}\vrule}l!{\color{black}\vrule}l!{\color{black}\vrule}l!{\color{black}\vrule}l!{\color{black}\vrule}l!{\color{black}\vrule}l!{\color{black}\vrule}l!{\color{black}\vrule}l!{\color{black}\vrule}l!{\color{black}\vrule}l!{\color{black}\vrule}l!{\color{black}\vrule}!{\color{black}\vrule}l!{\color{black}\vrule}l!{\color{black}\vrule}l!{\color{black}\vrule}l!{\color{black}\vrule}l!{\color{black}\vrule}l!{\color{black}\vrule}l!{\color{black}\vrule}l!{\color{black}\vrule}l!{\color{black}\vrule}l!{\color{black}\vrule}l!{\color{black}\vrule}l!{\color{black}\vrule}l!{\color{black}\vrule}!{\color{black}\vrule}l!{\color{black}\vrule}l!{\color{black}\vrule}l!{\color{black}\vrule}l!{\color{black}\vrule}l!{\color{black}\vrule}l!{\color{black}\vrule}l!{\color{black}\vrule}l!{\color{black}\vrule}l!{\color{black}\vrule}l!{\color{black}\vrule}l!{\color{black}\vrule}l!{\color{black}\vrule}l!{\color{black}\vrule}!{\color{black}\vrule}l!{\color{black}\vrule}l!{\color{black}\vrule}l!{\color{black}\vrule}l!{\color{black}\vrule}l!{\color{black}\vrule}l!{\color{black}\vrule}l!{\color{black}\vrule}l!{\color{black}\vrule}l!{\color{black}\vrule}l!{\color{black}\vrule}l!{\color{black}\vrule}l!{\color{black}\vrule}l!{\color{black}\vrule}} 
\hline
  %& \multicolumn{12}{l!{\color{black}\vrule}}{\textbf{MNIST-CNN}} 
  &\multicolumn{6}{c!{\color{black}\vrule}}{\textbf{Undefended CIFAR10-CNN}}
\\
\hline
\multirow{2}{*}{{\bf Baseline Attacks}}  & 
%\multicolumn{9}{l!{\color{black}\vrule}}{White-Box} &
%\multicolumn{3}{l!{\color{black}\vrule}}{Black-Box}&
\multicolumn{5}{c!{\color{black}\vrule}}{White-Box} 
&\multicolumn{1}{c!{\color{black}\vrule}}{Black-Box}

\\ 
%\hline
\cline{2-7}&  
 \multicolumn{2}{l!{\color{black}\vrule}}{\em FGS ($X_t=10K$)}   &  \multicolumn{2}{l!{\color{black}\vrule}}{\em PGD ($X_t=10K$)}   & \multicolumn{1}{l!{\color{black}\vrule}}{\em C\&W($X_t=5K$)} & \multicolumn{1}{l!{\color{black}\vrule}}{\em HSJA ($X_t=10K$)}   \\

\cline{1-7}
{\bf $||.||_p$ Norms}   & {\em $L_2$} & {\em $L_\infty$}    & {\em $L_2$}   & {\em $L_\infty$} & {\em $L_2$} & {\em $L_\infty$} \\
\hline\hline
\textbf{Initial Evasion Rate}    

& 22.36\%
& 83.04\%  
    
& 22.46\%         
& 91.07\%
& 99.22\%

& 14.04\%  
    \\
\hline
%\cline{1-9}
 \textbf{\sysname Evasion Rate}      
 
 & \textbf{30.14\%}
 & \textbf{87.45\%}  
   
 & \textbf{31.12\%}          
 & \textbf{92.34\%} 
 & 99.22\%
 
 & \textbf{28.87\%}  
   \\
\hline
%\cline{1-9}

%\cline{1-9}
\textbf{Average Perturbation Change}        

& \textbf{-48.52\%}
& \textbf{-49.62\%}  
   
&\textbf{-48.25\%}          
& \textbf{-51.15\%}  
& \textbf{-52.16\%} 

& \textbf{-36.18\%}
 \\
\hline
%\cline{1-9}
\end{tabular}}
\end{adjustbox}

 \vspace*{2em}
\caption{Summary of \sysname results on undefended CIFAR10-CNN Model ($\epsilon=0.3$).}
\label{tab:CIFAR-CNN}
\end{table*}

\begin{table*}[t]

\centering

\begin{adjustbox}{angle=0}
\scalebox{.99}{
\begin{tabular}{!{\color{black}\vrule}l!{\color{black}\vrule}l!{\color{black}\vrule}l!{\color{black}\vrule}l!{\color{black}\vrule}l!{\color{black}\vrule}l!{\color{black}\vrule}l!{\color{black}\vrule}l!{\color{black}\vrule}l!{\color{black}\vrule}l!{\color{black}\vrule}l!{\color{black}\vrule}l!{\color{black}\vrule}l!{\color{black}\vrule}!{\color{black}\vrule}l!{\color{black}\vrule}l!{\color{black}\vrule}l!{\color{black}\vrule}l!{\color{black}\vrule}l!{\color{black}\vrule}l!{\color{black}\vrule}l!{\color{black}\vrule}l!{\color{black}\vrule}l!{\color{black}\vrule}l!{\color{black}\vrule}l!{\color{black}\vrule}l!{\color{black}\vrule}l!{\color{black}\vrule}!{\color{black}\vrule}l!{\color{black}\vrule}l!{\color{black}\vrule}l!{\color{black}\vrule}l!{\color{black}\vrule}l!{\color{black}\vrule}l!{\color{black}\vrule}l!{\color{black}\vrule}l!{\color{black}\vrule}l!{\color{black}\vrule}l!{\color{black}\vrule}l!{\color{black}\vrule}l!{\color{black}\vrule}l!{\color{black}\vrule}!{\color{black}\vrule}l!{\color{black}\vrule}l!{\color{black}\vrule}l!{\color{black}\vrule}l!{\color{black}\vrule}l!{\color{black}\vrule}l!{\color{black}\vrule}l!{\color{black}\vrule}l!{\color{black}\vrule}l!{\color{black}\vrule}l!{\color{black}\vrule}l!{\color{black}\vrule}l!{\color{black}\vrule}l!{\color{black}\vrule}} 
\hline
  & 
\multicolumn{6}{c!{\color{black}\vrule}}{\textbf{Defended CIFAR10-ResNet}}
\\
\hline
\multirow{2}{*}{{\bf Baseline Attacks}}  & 
\multicolumn{5}{c!{\color{black}\vrule}}{White-Box} &
\multicolumn{1}{c!{\color{black}\vrule}}{Black-Box}

\\ 
%\hline
\cline{2-7}&  
  \multicolumn{2}{l!{\color{black}\vrule}}{\em FGS ($X_t=5K$)}   &  \multicolumn{2}{l!{\color{black}\vrule}}{\em PGD ($X_t=5K$)}   & \multicolumn{1}{l!{\color{black}\vrule}}{\em C\&W ($X_t=1K$)} & \multicolumn{1}{l!{\color{black}\vrule}}{\em SPSA ($X_t=1K$)}   \\

\cline{1-7}
{\bf $||.||_p$ Norms}  & {\em $L_2$} & {\em $L_\infty$}    & {\em $L_2$} & {\em $L_\infty$}   & {\em $L_2$}  & {\em $L_\infty$} \\
\hline\hline
\textbf{Initial Evasion Rate}

& 9.75\%        
& 73.27\%   

& 9.66\%
& 86.25\% 
    
& 99.00\%          
  
& 10.80\%
\\
\hline
%\cline{1-9}
 \textbf{\sysname Evasion Rate}

 & \textbf{18.05\%}
 & \textbf{74.73\%}

 & \textbf{18.37\%}
 & \textbf{86.76\%*}
 
 &\textbf{99.75\%}
 
 &\textbf{23.46\%}
\\
\hline
%\cline{1-9}
%\cline{1-9}
\textbf{Average Perturbation Change}           
&\textbf{-41.23\%}
&\textbf{-45.04\%}
&\textbf{-41.16\%}
&\textbf{-46.51\%}
&\textbf{-51.20\%}
&\textbf{-42.11\%}
\\
\hline
%\cline{1-9}

\end{tabular}}
\end{adjustbox}

\vspace*{2em}
\caption{Summary of \sysname results on defended (adversarially-trained) CIFAR10-ResNet50 Model ($\epsilon=0.3$).}
\label{tab:CIFAR-ResNet}
\end{table*}

\subsection{Evasion Results}\label{subsec:evasion-results}
Tables \ref{tab:MNIST-CNN}, \ref{tab:CIFAR-CNN}, and \ref{tab:CIFAR-ResNet} report results of \sysname on different networks; the undefended MINST-CNN, undefended CIFAR10-CNN, and defended (adversarially-trained) CIFAR10-ResNet, respectively.\\
\textbf{Evasion rate in a nutshell}: Across the three tables, we observe a significant increase in the evasion rate of studied baseline attacks after combining them with \sysname. For instance, Table \ref{tab:MNIST-CNN} shows an average increase of $28.78\%$ of the evasion rate across all attacks performed on the undefended MNIST-CNN model. Similarly, the evasion rate results in Table \ref{tab:CIFAR-CNN}, convey that, on average across all studied baseline attacks, $6.23\%$ more adversarial images are produced by \sysname to evade the undefended CIFAR10-CNN model. Same observations can be drawn from Table \ref{tab:CIFAR-ResNet}. More precisely, overall baseline attacks performed on the adversarially-trained CIFAR-ResNet model, we observe an average of $5.41\%$ increase in the evasion rate, when combined with \sysname. In a nutshell, our findings consistently suggest that explanation methods can be employed to effectively guide evasion attacks towards higher evasion accuracy.

\begin{figure*}[t!]
    
    \centering
    \scalebox{1}{
    %\framebox{
    \includegraphics[width=\textwidth]{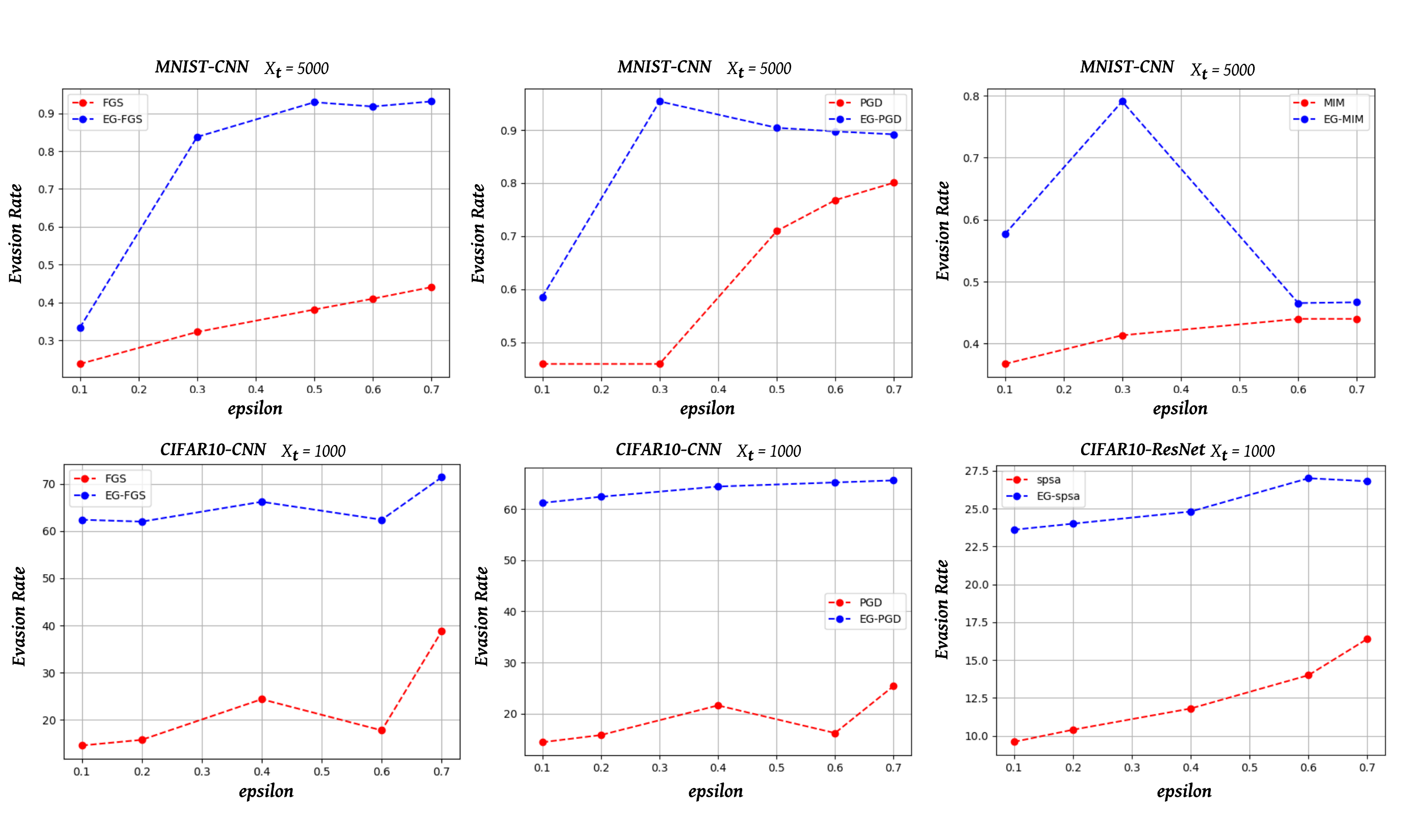}}
   % }
    \caption{The evasion rate curve of baseline attacks against \sysname attacks across different perturbation bounds $\epsilon$, using $||.||_\infty$ for MNIST and $||.||_2$ for CIFAR10 as distance metrics.}
    \label{fig:evasion-curve}
    % \vspace*{-1em}

\end{figure*}

\textbf{\sysname is consistently effective across model architectures, threat models, and distance metrics}: Our results consistently suggest that \sysname is agnostic to model architecture, threat model, and supports diverse distance metrics. For instance, the increase in evasion rate is observed on white-box baseline attacks (i.e., FGS, BIM, PGD, and C\&W) as well as for black-box attacks (i.e., MIM, HSJA, and SPSA). Additionally, the improvement in baseline attacks performance is observed for different $||.||_p$ norms. For the same perturbation bound value $\epsilon=0.3$ and the same attack strategy (e.g., FGS, PGD), we notice an increase in the evasion rate regardless of the employed $||.||_p$ distance metric.

\textbf{\sysname is consistently effective over a range of perturbation bounds}: Our experiments additionally cover the assessment of \sysname for different $\epsilon$ values. Figure \ref{fig:evasion-curve} reports the evasion rate curve of the state-of-the-art attacks (e.g., PGD, MIM,etc) before and after being guided with \sysname (see EG-PGD, EG-MIM, etc in Figure \ref{fig:evasion-curve}), using different perturbation bounds $\epsilon=0.1 \rightarrow 0.7$. For all target models trained on MNIST and CIFAR10, we observe a significant improvement in the evasion rate of all baseline attacks regardless of $\epsilon$. However, for most attacks, we observe that, a higher perturbation bound $\epsilon$ which allows a greater perturbation size can lead to a higher evasion rate.

It is noteworthy that the increase in rate of evasion results is considerably higher for baseline attacks that initially have a low evasion rate. For instance, \textit{FGS$_{L_\infty}$} and \textit{BIM$_{L_\infty}$} that initially produced, respectively, $28.20\%$ and $49.92\%$ evasive adversarial samples from the total of 10K MNIST samples, have improved at least twofold after employing \sysname. Similar observations can be drawn from CIFAR10-CNN and CIFAR10-ResNet. Particularly, for CIFAR10-CNN, the increase rate of evasion results on \textit{FGS$_{L_2}$}($22.36\% \rightarrow 30.14\%$) is higher than the increase rate of \textit{FGS$_{L_\infty}$} ($83.04\% \rightarrow 87.45\%$). However, even though \sysname results in higher increase rate in evasive samples for less preferment attacks, we note that, the total evasion rate of \sysname combined with stronger attacks is more important. This is expected, as \sysname builds on the initial results of baseline attacks which result in a correlation between the initial evasion rate of a baseline attack and the post-\sysname evasion rate. This is mainly observed for the C\&W attack, as across all models, combined with C\&W, \sysname exhibits the highest evasion rates (i.e., $>99\%$). Such correlation is also evident in the evasion rate curves from Figure \ref{fig:evasion-curve}.

\textbf{\sysname is still effective even against defended models}: In Table \ref{tab:CIFAR-ResNet}, we focus our experiments on a defended model in order to examine the effectiveness of \sysname against adversarial training defence. Since $PGD_{L_2}$ is used for adversarial training, CIFAR10-ResNet is expected to be specifically robust against gradient-based attacks performed using $L_2$ norm. Incontrovertibly, $FGS_{L_2}$ and $PGD_{L_2}$ achieve considerably lower initial evasion rates on CIFAR10-ResNet ($\sim 9\%$), compared to undefended models($\sim 20\%$). Nevertheless, combined with \sysname $FGS_{L_2}$ and $PGD_{L_2}$ result in more evasive samples ($\sim 18\%$), even against a defended model. Same findings can be observed for other baseline attacks (e.g. $FGS_{L_\infty}$, $SPSA_{L_\infty}$, etc). We conclude that \sysname is still effective even against defended models. Given this promising results, we hope that \sysname will be considered in the future as a benchmark for robustness assessment of ML models.

% Regrading the post-\sysname evasion results of baseline attacks, we conclude that ML feature-based explanations are useful to guide any $||.||_p$ norm-bounded attack towards making more consequential perturbations that lead to higher performance, not only against an undefended ML model, but also against an adversarially-trained model.\\

\subsection{Perturbation Change}\label{subsec:perturbation-change}
In addition to the evasion rate metric, we keep track of the perturbation changes introduced by \sysname for each experiment. More precisely, we compute the average of perturbation change including the number of added and eliminated perturbations (Equation \ref{eq:pert}). Results from the last rows of Tables \ref{tab:MNIST-CNN}, \ref{tab:CIFAR-CNN} and \ref{tab:CIFAR-ResNet} show that for most baseline attacks, \sysname is making less perturbations while improving the evasion rate. This is explained by the negative sign of the average perturbation change for most of the studied baseline attacks across the three tables. These findings prove that, without considering the pre-perturbation feature direction explanations, baseline attacks are initially performing a considerable number of non-consequential (unnecessary) perturbations. Consequently, in addition to the improvement of the evasion rate, these results demonstrate the importance of taking into account the features explanation weights in the formulation of evasion attacks (formulation \ref{eq:Booster}). 

It is noteworthy that the magnitude of the negative average perturbation change values is specifically more important for baseline attacks that initially have a high evasion rate. This is mainly true for the C\&W attack across the 3 tables and $PGD_{L_\infty}$ in Table \ref{tab:CIFAR-CNN}. We explain this observation by the fact that, \sysname performs additional perturbations only in the case that the baseline adversarial sample originally fails to evade the model (Algorithm \ref{alg:EG-Boost}:line 27), otherwise, it would only perform the elimination of non-consequential perturbations while maintaining the original evasion result.

In some of the experiments, we notice a positive value of the average perturbation change. This is particularly the case for $FGS_{L_\infty}$ and $BIM_{L_\infty}$ in Table \ref{tab:MNIST-CNN}. In this case, \sysname performs more additional perturbations than eliminations which reflects the drastic improvement of the evasion rate for these two attacks. These findings particularly demonstrate the direct impact of perturbing features that are directed to the true label in confusing the model to make a correct prediction.

Focusing on Tables \ref{tab:MNIST-CNN} and \ref{tab:CIFAR-CNN}, we notice that, on average, the increase rate in evasion results for MNIST ($28.78\%$) is higher than CIFAR-10 ($6.23\%$) using a CNN model for both datasets. Additionally, the average perturbation change for all experiments in Table \ref{tab:CIFAR-CNN} (i.e., CIFAR10-CNN) is negative and have a higher magnitude than their counterparts in Table \ref{tab:MNIST-CNN} (MNIST-CNN). Further investigations have shown that these two observations are related. In particular, we found that baseline attacks performed on CIFAR10-CNN are already perturbing a considerable number of features that are directed to the true label which makes the rate of added perturbations by the \sysname in CIFAR10-CNN lower than the ones added in MNIST-CNN. This observation might explain the difference in evasion increase between both datasets. However, as discussed in section \ref{sec: approach}, \sysname specifically examines this case i.e, the case where an important number of detected consequential features turn out to be already perturbed. More precisely, in case of initial evasion failure, \sysname proceeds by additionally perturbing features that are already perturbed by the baseline attack while ensuring that the perturbation size $\delta$ is still within the bound ($||\delta||_p<\epsilon$). 
% Experiments have shown that such a measure is especially useful for CIFAR10-CNN.

\begin{figure*}[t!]
    
    \centering
    % \scalebox{1}{
    %\framebox{
    \includegraphics[width=\textwidth]{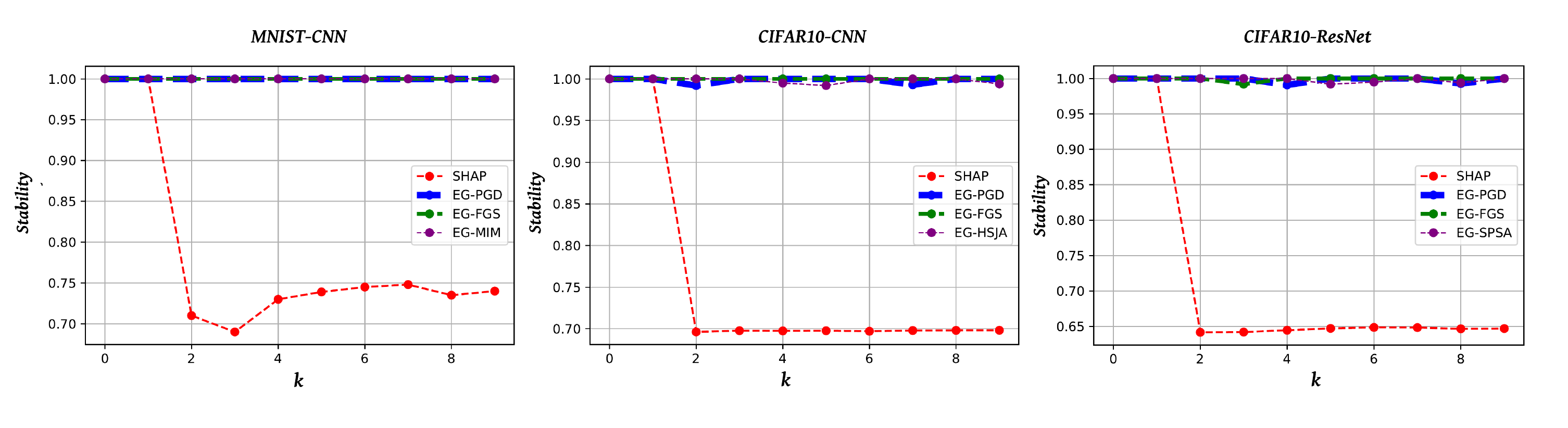}
   % }
    \caption{Comparison between the stability of \sysname and the employed explanation method, SHAP,  across models and datasets, for different $k$ values. $(k,l)-Stability$ of SHAP is computed on the top-10 features ($l=10$). \sysname is performed using different baseline attacks subject to $||\delta||_\infty < 0.3$. $k-$runs are performed on $1000$ test samples for each attack.}
    \label{fig:stability}
    % \vspace*{-1em}

\end{figure*}
\subsection{Stability Analysis}\label{subsec:stability-results}
As highlighted by prior works \cite{explainEval,explainEval2020} and thoroughly discussed in this paper, due to their stability concern, ML explanation methods should be carefully introduced when adopted in security-critical settings. Thus, we evaluate the reliability of explanation methods to be employed for the robustness assessment of ML systems by conducting a comparative analysis between the stability of the employed ML explainer (i.e., SHAP) and the stability of our explanation-based method. In particular, we use the \textit{(k-l)-Stability} and \textit{k-Stability} metrics defined in Section \ref{stab-ana}, to respectively compute the output's stability of SHAP, and \sysname combined with baseline attacks, across different studied models.

\textbf{\sysname doesn't inherit instabilities of SHAP}: In Figure \ref{fig:stability}, we plot our stability analysis results. Focusing on the stability curves of \sysname, we deduce that, it is almost $100\%$ stable for all studied models and across different baseline attacks. Consequently, the classification results returned by the target ML model are exactly the same across different runs of \sysname. Such findings prove the reliability of our approach to be confidently adopted for improved robustness assessment of ML models against baseline attacks. Compared to the stability curves of SHAP, we observe that \sysname does not inherit the instability concern indicated by ML explainer's output. As discussed in Section \ref{stab-ana}, these findings are explained by the reliance only on the sign of explanation weights to decide the feature directions (i.e.,\textit{positive} and \textit{negative} features). Since the distortions across different runs of the feature directions results returned by an accurate ML explainer are minor, the feature selection for perturbation performed by \sysname returns the same feature sets across different runs which overall leads to the same evasion results.

Although \sysname is not deeply influenced by the \textit{stability} of the employed ML explainer, its success is, however, relying on the their \textit{accuracy} to produce precise explanations, specifically precise feature directions in our case. Given the promising results, that our approach showed when relying on SHAP, we hope that future improvements in ML explanation methods would lead to an even better performance of \sysname.

\subsection{Execution Time}\label{subsec:exec-time}
The measurements we present here are based on a hardware environment with 6 vCPUs, 18.5 GB memory, and 1 GPU-NVIDIA Tesla K80.

\begin{figure*}[t!]
    
    \centering
    % \scalebox{0.52}{
    %\framebox{
    \includegraphics[width=\textwidth]{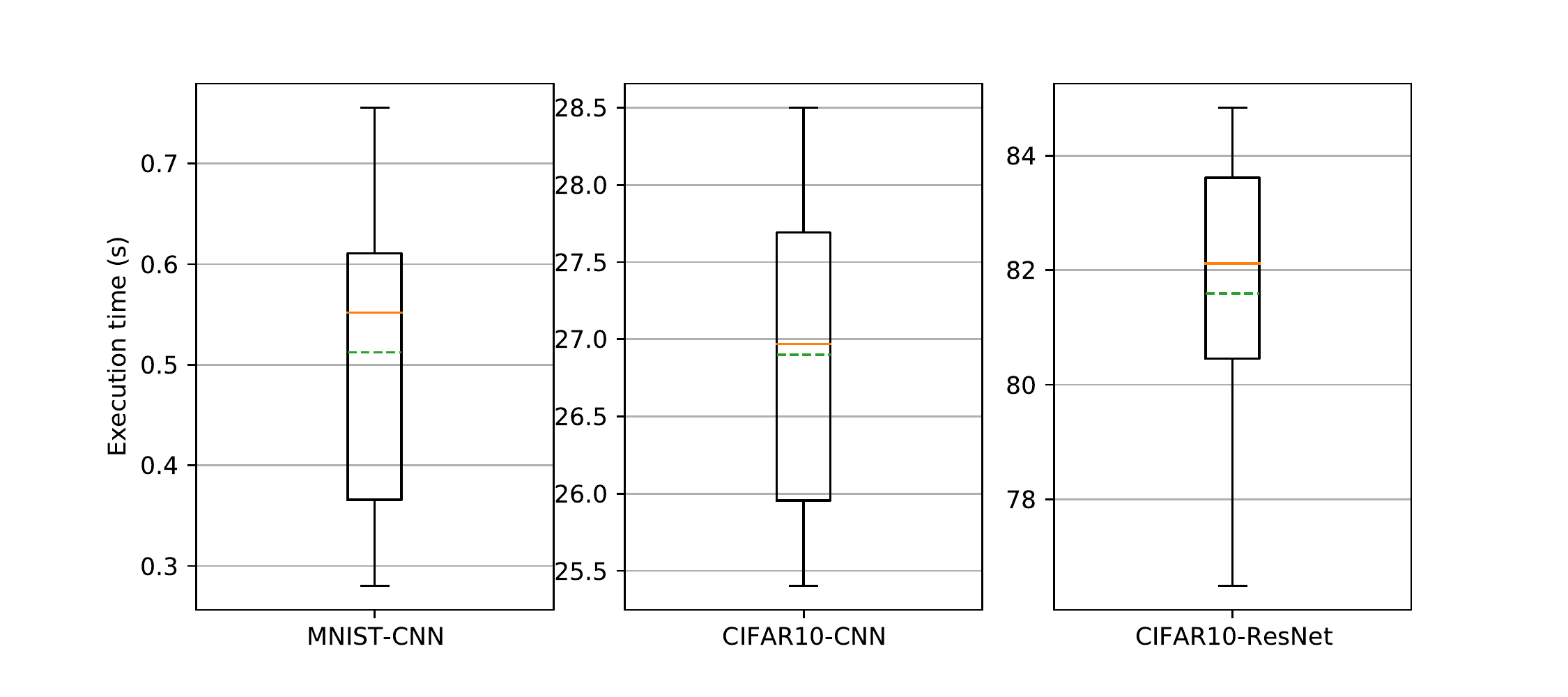}
   % }
    \caption{Execution time of \sysname performed on baseline attacks, per-sample, across the studied models. }
    \label{fig:time}
    % \vspace*{-1em}

\end{figure*}

In Figure \ref{fig:time}, we plot the range of execution times of \sysname recorded by different experiments. \sysname takes, on average  $0.55s$, $27.0s$ and $82.0s$ per-sample, respectively, for MNIST-CNN, CIFAR10-CNN, and CIFAR10-ResNet (orange line). Overall, these observations reflect the efficiency of \sysname, however, they additionally, reveal the impact of the input dimension and the model architecture on the running time of \sysname. For instance, it takes more time on CIFAR10, compared to MNIST, which is due to the difference in input dimension between both datasets i.e., (28,28) for MNIST and (3,32,32) for CIFAR10. Additionally, for the same dataset (i.e., CIFAR10), we observe a considerable difference in execution time of \sysname between CNN and ResNet50 (i.e., $27.0s \rightarrow 82.0s$) which is explained by the higher dimension of ResNet50 compared to CNN. This final result is expected, since \sysname is a query-based approach (Algorithm \ref{alg:EG-Boost}: lines 10,13,19, and 27). Thus, the duration that a neural network takes to respond to a query influences the total execution time of \sysname.

\section{Related Work}\label{sec: related}
In recent years, we have witnessed a flurry of adversarial example crafting methods~\cite{Carlini-list} and keeping up-to-date on new research in this sub-domain is a challenge. Nevertheless, in the following we position \sysname with respect to closely related work.\\

\textbf{Evasion Attacks.} Here, we focus on benchmark attacks that \sysname builds up on. While ML evasion attacks can be traced back to early 2000s~\cite{Wild-patterns18}, recent improvements in accuracy of deep neural networks on non-trivial learning tasks have triggered  adversarial manipulations as potential threats to ML models. Biggio et al.~\cite{Biggio-ECML13} introduced gradient-based evasion attacks against PDF malware detectors. Szegedy et al.~\cite{szegedy2014intriguing} demonstrated gradient-based attacks against image classifiers. Goodfellow et al.~\cite{FGSM} introduced FGSM, a fast one-step adversarial example generation strategy. Kurakin et al.~\cite{BIM} later introduced BIM as an iterative improvement of FGSM. Around the same time, Madry et al.~\cite{PGSM} proposed PGSM and Uesato et al.~\cite{PGSM} later introduced PGD. The common thread among these class of attacks and similar gradient-based ones is that they all aim to solve for an adversarial example misclassified with maximum confidence subject to a bounded perturbation size.

Another line of evasion attacks aim to minimize the perturbation distance subject to the evasion goal (e.g., classify a sample to a target label chosen by the adversary). Among such attacks, Carlini and Wagner~\cite{CW} introduced the C\&W attack, one of the strongest white-box attacks to date. Other attacks such as DeepFool~\cite{DeepFool16} and SparseFool~\cite{SparseFool19} use the same notion of minimum perturbation distance. Recent attacks introduced by Brendel et al.~\cite{Accurate-Fast19} and Croce and Hein~\cite{Croce020a} improve on the likes of DeepFool and SparseFool, both on accuracy and speed of adversarial example generation. {\em \sysname builds up on this body of adversarial example crafting attacks. As our evaluations suggest, it improves the evasion rate of major benchmark attacks used by prior work}. 

Pintor et  al.~\cite{FMN21} propose the fast minimum-norm (FMN) attack that improves both accuracy and speed of benchmark gradient-based attacks across $L_p$ norms ($l = 0,1,2,\infty$), for both targeted and untargeted white-box attacks. Like FMN, \sysname covers MNIST and CIFAR10 datasets, evaluated  on undefended and defended models, considers $L_p$ norm distance metrics, and  on aggregate improves evasion accuracy of  existing attacks. Unlike FMN, \sysname covers black-box and  white-box attacks, is not limited to gradient-based attacks, instead of proposing an attack strategy it rather leverages explanations to improve accuracy of existing attacks, and is limited to untargeted attacks. \\

Among black-box evasion attacks, we highlight MIM~\cite{MIM}, HSJA~\cite{HSJA20}, and SPSA~\cite{uesato2018adversarial}. MIM  leverages the momentum method~\cite{momentum} that stabilizes gradient updates by accumulating a velocity vector in the gradient direction of the loss function across iterations, and allows the algorithm to escape from poor local maxima. HSJA  relies on prediction label only to create adversarial examples, with key intuition based on gradient-direction estimation, motivated by zeroth-order optimization. In SPSA, Uesato et al.~\cite{uesato2018adversarial} explore adversarial risk as a measure of the model’s performance on worst-case inputs, to define a tractable surrogate objective to the true adversarial risk.\\

\textbf{Explanation-Guided Attacks.} A recent body of work demonstrate how ML explanation methods can be harnessed to poison training data, extract model, and infer training data-points. Severi et al.~\cite{EG-Poisoning} leverage ML explanation methods to guide the selection of features to construct backdoor triggers that poison training data of malware classifiers. Milli et al.~\cite{EG-model-extraction19} demonstrate that gradient-based model explanations can be leveraged to approximate the underlying model. In a similar line of work, Aïvodji. et al.~\cite{EG-model-extraction20} show how counterfactual explanations can be used to extract a model. Very recently, Shokri et al.~\cite{EG-mem-inference2021} study the tension between transparency via model explanations and privacy leakage via membership inference, and show backpropagation-based explanations can leak a significant amount of information about individual training data-points. {\em Compared to this body of work, \sysname is the first approach to leverage ML explanations to significantly improve accuracy of evasion attacks}.

\textbf{Reliability Analysis of Explanation Methods.}
Focusing on robustness of explanations to adversarial manipulations, Ghorbani et al.~\cite{exp-Rob1} and  Heo et al.~\cite{exp-Rob2} demonstrate that model explanation results are sensitive to small perturbations of explanations that preserve the predicted label. If ML explanations were provided to \sysname as in ``Explanation-as-a-Service'', an adversary could potentially alter the explanation results, which might in effect influence our explanation-guided attack. {\em In \sysname, baseline attacks and explanation methods are locally integrated into the robustness evaluation workflow (hence not disclosed to an adversary)}.

Recent studies by Warnecke et al.~\cite{explainEval} and Fan et al.~\cite{explainEval2020} systematically analyze the utility of ML explanation methods especially on models trained for security-critical tasks (e.g., malware classifiers). In addition to general evaluation criteria (e.g., explanation accuracy and sparsity), Warnecke et al.~\cite{explainEval} focused on other security-relevant evaluation metrics (e.g., stability, efficiency, and robustness). In a related line of work, Fan et al.~\cite{explainEval2020} performed a similar study on Android malware classifiers that led to similar conclusions.
\section{Conclusion}\label{sec: concl}
In this paper, we introduce \sysname, the first explanation-guided booster for ML evasion attacks. Guided by feature-based explanations of model predictions, \sysname significantly improves evasion accuracy of state-of-the-art adversarial example crafting attacks by introducing consequential perturbations and eliminating non-consequential ones. By systematically evaluating \sysname on MNIST and CIFAR10, across white-box and black-box attacks against undefended and defended models, and using diverse distance metrics, we show that \sysname significantly improves evasion accuracy of reference evasion attacks on undefended MNIST-CNN and CIFAR10-CNN models, and an adversarially-trained CIFAR10-ResNet50 model. Furthermore, we empirically show the reliability of explanation methods to be adopted for robustness assessment of ML models. We hope that \sysname will be used by future work as a benchmark for robustness evaluation of ML models against adversarial examples.
\bibliographystyle{ACM-Reference-Format}
\bibliography{main}


\begin{thebibliography}{43}

%%% the Web address.

\ifx \showCODEN    \undefined \def \showCODEN     #1{\unskip}     \fi
\ifx \showDOI      \undefined \def \showDOI       #1{#1}\fi
\ifx \showISBNx    \undefined \def \showISBNx     #1{\unskip}     \fi
\ifx \showISBNxiii \undefined \def \showISBNxiii  #1{\unskip}     \fi
\ifx \showISSN     \undefined \def \showISSN      #1{\unskip}     \fi
\ifx \showLCCN     \undefined \def \showLCCN      #1{\unskip}     \fi
\ifx \shownote     \undefined \def \shownote      #1{#1}          \fi
\ifx \showarticletitle \undefined \def \showarticletitle #1{#1}   \fi
\ifx \showURL      \undefined \def \showURL       {\relax}        \fi
% invisible to TeX
\providecommand\bibfield[2]{#2}
\providecommand\bibinfo[2]{#2}
\providecommand\natexlab[1]{#1}
\providecommand\showeprint[2][]{arXiv:#2}

\bibitem[\protect\citeauthoryear{??}{CIF}{2020}]%
        {CIFAR-CNN}
 \bibinfo{year}{2020}\natexlab{}.
\newblock \bibinfo{title}{{CNN-CIFAR10 model}}.
\newblock
  \bibinfo{howpublished}{\url{https://github.com/jamespengcheng/PyTorch-CNN-on-CIFAR10}}.
\newblock


\bibitem[\protect\citeauthoryear{Amich and Eshete}{Amich and Eshete}{2021}]%
        {amich2021explanationguided}
\bibfield{author}{\bibinfo{person}{Abderrahmen Amich} {and}
  \bibinfo{person}{Birhanu Eshete}.} \bibinfo{year}{2021}\natexlab{}.
\newblock \showarticletitle{{Explanation-Guided Diagnosis of Machine Learning
  Evasion Attacks}}. In \bibinfo{booktitle}{\emph{Security and Privacy in
  Communication Networks - 17th {EAI} International Conference, SecureComm
  2021}}.
\newblock


\bibitem[\protect\citeauthoryear{Andor, Alberti, Weiss, Severyn, Presta,
  Ganchev, Petrov, and Collins}{Andor et~al\mbox{.}}{2016}]%
        {NLP-16}
\bibfield{author}{\bibinfo{person}{Daniel Andor}, \bibinfo{person}{Chris
  Alberti}, \bibinfo{person}{David Weiss}, \bibinfo{person}{Aliaksei Severyn},
  \bibinfo{person}{Alessandro Presta}, \bibinfo{person}{Kuzman Ganchev},
  \bibinfo{person}{Slav Petrov}, {and} \bibinfo{person}{Michael Collins}.}
  \bibinfo{year}{2016}\natexlab{}.
\newblock \showarticletitle{Globally Normalized Transition-Based Neural
  Networks}. In \bibinfo{booktitle}{\emph{Proceedings of the 54th Annual
  Meeting of the Association for Computational Linguistics, {ACL} 2016, August
  7-12, 2016, Berlin, Germany, Volume 1: Long Papers}}. \bibinfo{publisher}{The
  Association for Computer Linguistics}.
\newblock


\bibitem[\protect\citeauthoryear{Aïvodji, Bolot, and Gambs}{Aïvodji
  et~al\mbox{.}}{2020}]%
        {EG-model-extraction20}
\bibfield{author}{\bibinfo{person}{Ulrich Aïvodji}, \bibinfo{person}{Alexandre
  Bolot}, {and} \bibinfo{person}{Sébastien Gambs}.}
  \bibinfo{year}{2020}\natexlab{}.
\newblock \bibinfo{title}{Model extraction from counterfactual explanations}.
\newblock
\newblock
\showeprint[arxiv]{2009.01884}~[cs.LG]


\bibitem[\protect\citeauthoryear{Biggio, Corona, Maiorca, Nelson, Srndic,
  Laskov, Giacinto, and Roli}{Biggio et~al\mbox{.}}{2013}]%
        {Biggio-ECML13}
\bibfield{author}{\bibinfo{person}{Battista Biggio}, \bibinfo{person}{Igino
  Corona}, \bibinfo{person}{Davide Maiorca}, \bibinfo{person}{Blaine Nelson},
  \bibinfo{person}{Nedim Srndic}, \bibinfo{person}{Pavel Laskov},
  \bibinfo{person}{Giorgio Giacinto}, {and} \bibinfo{person}{Fabio Roli}.}
  \bibinfo{year}{2013}\natexlab{}.
\newblock \showarticletitle{Evasion Attacks against Machine Learning at Test
  Time}. In \bibinfo{booktitle}{\emph{Machine Learning and Knowledge Discovery
  in Databases - European Conference, {ECML} {PKDD} 2013, Prague, Czech
  Republic, September 23-27, 2013, Proceedings, Part {III}}}.
  \bibinfo{pages}{387--402}.
\newblock


\bibitem[\protect\citeauthoryear{Biggio and Roli}{Biggio and Roli}{2018}]%
        {Wild-patterns18}
\bibfield{author}{\bibinfo{person}{Battista Biggio} {and}
  \bibinfo{person}{Fabio Roli}.} \bibinfo{year}{2018}\natexlab{}.
\newblock \showarticletitle{Wild patterns: Ten years after the rise of
  adversarial machine learning}.
\newblock \bibinfo{journal}{\emph{Pattern Recognition}}  \bibinfo{volume}{84}
  (\bibinfo{year}{2018}), \bibinfo{pages}{317--331}.
\newblock


\bibitem[\protect\citeauthoryear{Brendel, Rauber, K{\"{u}}mmerer,
  Ustyuzhaninov, and Bethge}{Brendel et~al\mbox{.}}{2019}]%
        {Accurate-Fast19}
\bibfield{author}{\bibinfo{person}{Wieland Brendel}, \bibinfo{person}{Jonas
  Rauber}, \bibinfo{person}{Matthias K{\"{u}}mmerer}, \bibinfo{person}{Ivan
  Ustyuzhaninov}, {and} \bibinfo{person}{Matthias Bethge}.}
  \bibinfo{year}{2019}\natexlab{}.
\newblock \showarticletitle{Accurate, reliable and fast robustness evaluation}.
  In \bibinfo{booktitle}{\emph{Advances in Neural Information Processing
  Systems 32: Annual Conference on Neural Information Processing Systems 2019,
  NeurIPS 2019, December 8-14, 2019, Vancouver, BC, Canada}}.
  \bibinfo{pages}{12841--12851}.
\newblock


\bibitem[\protect\citeauthoryear{Carlini}{Carlini}{2020}]%
        {Carlini-list}
\bibfield{author}{\bibinfo{person}{Nicholas Carlini}.}
  \bibinfo{year}{2020}\natexlab{}.
\newblock \bibinfo{title}{{A Complete List of All (arXiv) Adversarial Example
  Papers}}.
\newblock
  \bibinfo{howpublished}{\url{https://nicholas.carlini.com/writing/2019/all-adversarial-example-papers.html}}.
\newblock


\bibitem[\protect\citeauthoryear{Carlini and Wagner}{Carlini and
  Wagner}{2017}]%
        {CW}
\bibfield{author}{\bibinfo{person}{Nicholas Carlini} {and}
  \bibinfo{person}{David~A. Wagner}.} \bibinfo{year}{2017}\natexlab{}.
\newblock \showarticletitle{Towards Evaluating the Robustness of Neural
  Networks}. In \bibinfo{booktitle}{\emph{2017 {IEEE} Symposium on Security and
  Privacy, {SP} 2017, San Jose, CA, USA, May 22-26, 2017}}.
  \bibinfo{pages}{39--57}.
\newblock


\bibitem[\protect\citeauthoryear{Chen, Jordan, and Wainwright}{Chen
  et~al\mbox{.}}{2020}]%
        {HSJA20}
\bibfield{author}{\bibinfo{person}{Jianbo Chen}, \bibinfo{person}{Michael~I.
  Jordan}, {and} \bibinfo{person}{Martin~J. Wainwright}.}
  \bibinfo{year}{2020}\natexlab{}.
\newblock \bibinfo{title}{HopSkipJumpAttack: A Query-Efficient Decision-Based
  Attack}.
\newblock
\newblock
\showeprint[arxiv]{1904.02144}~[cs.LG]


\bibitem[\protect\citeauthoryear{Croce and Hein}{Croce and Hein}{2020}]%
        {Croce020a}
\bibfield{author}{\bibinfo{person}{Francesco Croce} {and}
  \bibinfo{person}{Matthias Hein}.} \bibinfo{year}{2020}\natexlab{}.
\newblock \showarticletitle{Reliable evaluation of adversarial robustness with
  an ensemble of diverse parameter-free attacks}. In
  \bibinfo{booktitle}{\emph{Proceedings of the 37th International Conference on
  Machine Learning, {ICML} 2020, 13-18 July 2020, Virtual Event}}
  \emph{(\bibinfo{series}{Proceedings of Machine Learning Research},
  Vol.~\bibinfo{volume}{119})}. \bibinfo{publisher}{{PMLR}},
  \bibinfo{pages}{2206--2216}.
\newblock


\bibitem[\protect\citeauthoryear{Dahl, Yu, Deng, and Acero}{Dahl
  et~al\mbox{.}}{2012}]%
        {DL-Speech2012}
\bibfield{author}{\bibinfo{person}{G.~E. Dahl}, \bibinfo{person}{Dong Yu},
  \bibinfo{person}{Li Deng}, {and} \bibinfo{person}{A. Acero}.}
  \bibinfo{year}{2012}\natexlab{}.
\newblock \showarticletitle{Context-Dependent Pre-Trained Deep Neural Networks
  for Large-Vocabulary Speech Recognition}.
\newblock \bibinfo{journal}{\emph{{IEEE} Transactions on Audio, Speech, and
  Language Processing}} \bibinfo{volume}{20}, \bibinfo{number}{1}
  (\bibinfo{date}{Jan.} \bibinfo{year}{2012}), \bibinfo{pages}{30--42}.
\newblock


\bibitem[\protect\citeauthoryear{Dong, Liao, Pang, Su, Zhu, Hu, and Li}{Dong
  et~al\mbox{.}}{2018}]%
        {MIM}
\bibfield{author}{\bibinfo{person}{Yinpeng Dong}, \bibinfo{person}{Fangzhou
  Liao}, \bibinfo{person}{Tianyu Pang}, \bibinfo{person}{Hang Su},
  \bibinfo{person}{Jun Zhu}, \bibinfo{person}{Xiaolin Hu}, {and}
  \bibinfo{person}{Jianguo Li}.} \bibinfo{year}{2018}\natexlab{}.
\newblock \showarticletitle{Boosting Adversarial Attacks With Momentum}. In
  \bibinfo{booktitle}{\emph{2018 {IEEE} Conference on Computer Vision and
  Pattern Recognition, {CVPR} 2018, Salt Lake City, UT, USA, June 18-22,
  2018}}. \bibinfo{publisher}{{IEEE} Computer Society},
  \bibinfo{pages}{9185--9193}.
\newblock


\bibitem[\protect\citeauthoryear{Engstrom, Ilyas, Salman, Santurkar, and
  Tsipras}{Engstrom et~al\mbox{.}}{2019}]%
        {robustness}
\bibfield{author}{\bibinfo{person}{Logan Engstrom}, \bibinfo{person}{Andrew
  Ilyas}, \bibinfo{person}{Hadi Salman}, \bibinfo{person}{Shibani Santurkar},
  {and} \bibinfo{person}{Dimitris Tsipras}.} \bibinfo{year}{2019}\natexlab{}.
\newblock \bibinfo{title}{Robustness (Python Library)}.
\newblock
\newblock
\urldef\tempurl%
\url{https://github.com/MadryLab/robustness}
\showURL{%
\tempurl}


\bibitem[\protect\citeauthoryear{{Fan}, {Wei}, {Xie}, {Liu}, {Guan}, and
  {Liu}}{{Fan} et~al\mbox{.}}{2021}]%
        {explainEval2020}
\bibfield{author}{\bibinfo{person}{M. {Fan}}, \bibinfo{person}{W. {Wei}},
  \bibinfo{person}{X. {Xie}}, \bibinfo{person}{Y. {Liu}}, \bibinfo{person}{X.
  {Guan}}, {and} \bibinfo{person}{T. {Liu}}.} \bibinfo{year}{2021}\natexlab{}.
\newblock \showarticletitle{Can We Trust Your Explanations? Sanity Checks for
  Interpreters in Android Malware Analysis}.
\newblock \bibinfo{journal}{\emph{IEEE Transactions on Information Forensics
  and Security}}  \bibinfo{volume}{16} (\bibinfo{year}{2021}),
  \bibinfo{pages}{838--853}.
\newblock


\bibitem[\protect\citeauthoryear{Ghorbani, Abid, and Zou}{Ghorbani
  et~al\mbox{.}}{2017}]%
        {exp-Rob1}
\bibfield{author}{\bibinfo{person}{Amirata Ghorbani}, \bibinfo{person}{Abubakar
  Abid}, {and} \bibinfo{person}{James Zou}.} \bibinfo{year}{2017}\natexlab{}.
\newblock \showarticletitle{Interpretation of Neural Networks Is Fragile}.
\newblock \bibinfo{journal}{\emph{Proceedings of the AAAI Conference on
  Artificial Intelligence}}  \bibinfo{volume}{33} (\bibinfo{date}{10}
  \bibinfo{year}{2017}).
\newblock


\bibitem[\protect\citeauthoryear{Goodfellow, Shlens, and Szegedy}{Goodfellow
  et~al\mbox{.}}{2015}]%
        {FGSM}
\bibfield{author}{\bibinfo{person}{Ian~J. Goodfellow},
  \bibinfo{person}{Jonathon Shlens}, {and} \bibinfo{person}{Christian
  Szegedy}.} \bibinfo{year}{2015}\natexlab{}.
\newblock \showarticletitle{Explaining and Harnessing Adversarial Examples}. In
  \bibinfo{booktitle}{\emph{3rd International Conference on Learning
  Representations, {ICLR} 2015, San Diego, CA, USA, May 7-9, 2015, Conference
  Track Proceedings}}.
\newblock


\bibitem[\protect\citeauthoryear{Guo, Mu, Xu, Su, Wang, and Xing}{Guo
  et~al\mbox{.}}{2018}]%
        {LEMNA}
\bibfield{author}{\bibinfo{person}{Wenbo Guo}, \bibinfo{person}{Dongliang Mu},
  \bibinfo{person}{Jun Xu}, \bibinfo{person}{Purui Su}, \bibinfo{person}{Gang
  Wang}, {and} \bibinfo{person}{Xinyu Xing}.} \bibinfo{year}{2018}\natexlab{}.
\newblock \showarticletitle{{LEMNA:} Explaining Deep Learning based Security
  Applications}. In \bibinfo{booktitle}{\emph{Proceedings of the 2018 {ACM}
  {SIGSAC} Conference on Computer and Communications Security, {CCS} 2018,
  Toronto, ON, Canada, October 15-19, 2018}}. \bibinfo{publisher}{{ACM}},
  \bibinfo{pages}{364--379}.
\newblock


\bibitem[\protect\citeauthoryear{Heo, Joo, and Moon}{Heo et~al\mbox{.}}{2019}]%
        {exp-Rob2}
\bibfield{author}{\bibinfo{person}{Juyeon Heo}, \bibinfo{person}{Sunghwan Joo},
  {and} \bibinfo{person}{Taesup Moon}.} \bibinfo{year}{2019}\natexlab{}.
\newblock \showarticletitle{Fooling Neural Network Interpretations via
  Adversarial Model Manipulation}.
\newblock  (\bibinfo{date}{02} \bibinfo{year}{2019}).
\newblock


\bibitem[\protect\citeauthoryear{Kolosnjaji, Demontis, Biggio, Maiorca,
  Giacinto, Eckert, and Roli}{Kolosnjaji et~al\mbox{.}}{2018}]%
        {MalConvEvade18}
\bibfield{author}{\bibinfo{person}{Bojan Kolosnjaji}, \bibinfo{person}{Ambra
  Demontis}, \bibinfo{person}{Battista Biggio}, \bibinfo{person}{Davide
  Maiorca}, \bibinfo{person}{Giorgio Giacinto}, \bibinfo{person}{Claudia
  Eckert}, {and} \bibinfo{person}{Fabio Roli}.}
  \bibinfo{year}{2018}\natexlab{}.
\newblock \showarticletitle{Adversarial Malware Binaries: Evading Deep Learning
  for Malware Detection in Executables}. In \bibinfo{booktitle}{\emph{26th
  European Signal Processing Conference, {EUSIPCO} 2018, Roma, Italy, September
  3-7, 2018}}. \bibinfo{pages}{533--537}.
\newblock


\bibitem[\protect\citeauthoryear{Krizhevsky, Nair, and Hinton}{Krizhevsky
  et~al\mbox{.}}{[n.d.]}]%
        {cifar}
\bibfield{author}{\bibinfo{person}{Alex Krizhevsky}, \bibinfo{person}{Vinod
  Nair}, {and} \bibinfo{person}{Geoffrey Hinton}.}
  \bibinfo{year}{[n.d.]}\natexlab{}.
\newblock \showarticletitle{CIFAR-10 (Canadian Institute for Advanced
  Research)}.
\newblock  (\bibinfo{year}{[n.\,d.]}).
\newblock
\urldef\tempurl%
\url{http://www.cs.toronto.edu/~kriz/cifar.html}
\showURL{%
\tempurl}


\bibitem[\protect\citeauthoryear{Kurakin, Goodfellow, and Bengio}{Kurakin
  et~al\mbox{.}}{2016}]%
        {BIM}
\bibfield{author}{\bibinfo{person}{Alexey Kurakin}, \bibinfo{person}{Ian~J.
  Goodfellow}, {and} \bibinfo{person}{Samy Bengio}.}
  \bibinfo{year}{2016}\natexlab{}.
\newblock \showarticletitle{Adversarial Machine Learning at Scale}.
\newblock \bibinfo{journal}{\emph{CoRR}}  \bibinfo{volume}{abs/1611.01236}
  (\bibinfo{year}{2016}).
\newblock
\showeprint[arxiv]{1611.01236}


\bibitem[\protect\citeauthoryear{LeCun, Cortes, and Burges}{LeCun
  et~al\mbox{.}}{2020}]%
        {MNIST}
\bibfield{author}{\bibinfo{person}{Yan LeCun}, \bibinfo{person}{Corinna
  Cortes}, {and} \bibinfo{person}{Christopher~J.C. Burges}.}
  \bibinfo{year}{2020}\natexlab{}.
\newblock \bibinfo{title}{The MNIST Database of Handwritten Digits}.
\newblock \bibinfo{howpublished}{\url{http://yann.lecun.com/exdb/mnist/}}.
\newblock


\bibitem[\protect\citeauthoryear{Lundberg and Lee}{Lundberg and Lee}{2017}]%
        {SHAP}
\bibfield{author}{\bibinfo{person}{Scott~M. Lundberg} {and}
  \bibinfo{person}{Su{-}In Lee}.} \bibinfo{year}{2017}\natexlab{}.
\newblock \showarticletitle{A Unified Approach to Interpreting Model
  Predictions}. In \bibinfo{booktitle}{\emph{Advances in Neural Information
  Processing Systems 30: Annual Conference on Neural Information Processing
  Systems 2017, 4-9 December 2017, Long Beach, CA, {USA}}}.
  \bibinfo{pages}{4765--4774}.
\newblock


\bibitem[\protect\citeauthoryear{Madry, Makelov, Schmidt, Tsipras, and
  Vladu}{Madry et~al\mbox{.}}{2017}]%
        {PGSM}
\bibfield{author}{\bibinfo{person}{Aleksander Madry},
  \bibinfo{person}{Aleksandar Makelov}, \bibinfo{person}{Ludwig Schmidt},
  \bibinfo{person}{Dimitris Tsipras}, {and} \bibinfo{person}{Adrian Vladu}.}
  \bibinfo{year}{2017}\natexlab{}.
\newblock \showarticletitle{Towards Deep Learning Models Resistant to
  Adversarial Attacks}.
\newblock \bibinfo{journal}{\emph{CoRR}}  \bibinfo{volume}{abs/1706.06083}
  (\bibinfo{year}{2017}).
\newblock
\showeprint{1706.06083}


\bibitem[\protect\citeauthoryear{Melis, Demontis, Pintor, Sotgiu, and
  Biggio}{Melis et~al\mbox{.}}{2019}]%
        {melis2019secml}
\bibfield{author}{\bibinfo{person}{Marco Melis}, \bibinfo{person}{Ambra
  Demontis}, \bibinfo{person}{Maura Pintor}, \bibinfo{person}{Angelo Sotgiu},
  {and} \bibinfo{person}{Battista Biggio}.} \bibinfo{year}{2019}\natexlab{}.
\newblock \showarticletitle{secml: A Python Library for Secure and Explainable
  Machine Learning}.
\newblock \bibinfo{journal}{\emph{arXiv preprint arXiv:1912.10013}}
  (\bibinfo{year}{2019}).
\newblock


\bibitem[\protect\citeauthoryear{Milli, Schmidt, Dragan, and Hardt}{Milli
  et~al\mbox{.}}{2019}]%
        {EG-model-extraction19}
\bibfield{author}{\bibinfo{person}{Smitha Milli}, \bibinfo{person}{Ludwig
  Schmidt}, \bibinfo{person}{Anca~D. Dragan}, {and} \bibinfo{person}{Moritz
  Hardt}.} \bibinfo{year}{2019}\natexlab{}.
\newblock \showarticletitle{Model Reconstruction from Model Explanations}. In
  \bibinfo{booktitle}{\emph{Proceedings of the Conference on Fairness,
  Accountability, and Transparency, FAT* 2019, Atlanta, GA, USA, January 29-31,
  2019}}. \bibinfo{publisher}{{ACM}}, \bibinfo{pages}{1--9}.
\newblock


\bibitem[\protect\citeauthoryear{Modas, Moosavi{-}Dezfooli, and Frossard}{Modas
  et~al\mbox{.}}{2019}]%
        {SparseFool19}
\bibfield{author}{\bibinfo{person}{Apostolos Modas},
  \bibinfo{person}{Seyed{-}Mohsen Moosavi{-}Dezfooli}, {and}
  \bibinfo{person}{Pascal Frossard}.} \bibinfo{year}{2019}\natexlab{}.
\newblock \showarticletitle{SparseFool: {A} Few Pixels Make a Big Difference}.
  In \bibinfo{booktitle}{\emph{{IEEE} Conference on Computer Vision and Pattern
  Recognition, {CVPR} 2019, Long Beach, CA, USA, June 16-20, 2019}}.
  \bibinfo{publisher}{Computer Vision Foundation / {IEEE}},
  \bibinfo{pages}{9087--9096}.
\newblock


\bibitem[\protect\citeauthoryear{Moosavi{-}Dezfooli, Fawzi, and
  Frossard}{Moosavi{-}Dezfooli et~al\mbox{.}}{2016}]%
        {DeepFool16}
\bibfield{author}{\bibinfo{person}{Seyed{-}Mohsen Moosavi{-}Dezfooli},
  \bibinfo{person}{Alhussein Fawzi}, {and} \bibinfo{person}{Pascal Frossard}.}
  \bibinfo{year}{2016}\natexlab{}.
\newblock \showarticletitle{DeepFool: {A} Simple and Accurate Method to Fool
  Deep Neural Networks}. In \bibinfo{booktitle}{\emph{2016 {IEEE} Conference on
  Computer Vision and Pattern Recognition, {CVPR} 2016, Las Vegas, NV, USA,
  June 27-30, 2016}}. \bibinfo{publisher}{{IEEE} Computer Society},
  \bibinfo{pages}{2574--2582}.
\newblock


\bibitem[\protect\citeauthoryear{Papernot, McDaniel, Wu, Jha, and
  Swami}{Papernot et~al\mbox{.}}{2016}]%
        {distillation}
\bibfield{author}{\bibinfo{person}{Nicolas Papernot}, \bibinfo{person}{Patrick
  McDaniel}, \bibinfo{person}{Xi Wu}, \bibinfo{person}{Somesh Jha}, {and}
  \bibinfo{person}{Ananthram Swami}.} \bibinfo{year}{2016}\natexlab{}.
\newblock \showarticletitle{Distillation as a Defense to Adversarial
  Perturbations Against Deep Neural Networks}. \bibinfo{pages}{582--597}.
\newblock


\bibitem[\protect\citeauthoryear{Pintor, Roli, Brendel, and Biggio}{Pintor
  et~al\mbox{.}}{2021}]%
        {FMN21}
\bibfield{author}{\bibinfo{person}{Maura Pintor}, \bibinfo{person}{Fabio Roli},
  \bibinfo{person}{Wieland Brendel}, {and} \bibinfo{person}{Battista Biggio}.}
  \bibinfo{year}{2021}\natexlab{}.
\newblock \showarticletitle{Fast Minimum-norm Adversarial Attacks through
  Adaptive Norm Constraints}.
\newblock \bibinfo{journal}{\emph{CoRR}}  \bibinfo{volume}{abs/2102.12827}
  (\bibinfo{year}{2021}).
\newblock


\bibitem[\protect\citeauthoryear{Polyak}{Polyak}{1964}]%
        {momentum}
\bibfield{author}{\bibinfo{person}{Boris~T. Polyak}.}
  \bibinfo{year}{1964}\natexlab{}.
\newblock \showarticletitle{Some Methods of Speeding up the Convergence of
  Iteration Methods}.
\newblock \bibinfo{journal}{\emph{U. S. S. R. Comput. Math. and Math. Phys.}}
  \bibinfo{volume}{4} (\bibinfo{year}{1964}), \bibinfo{pages}{1--17}.
\newblock
Issue 5.


\bibitem[\protect\citeauthoryear{Ribeiro, Singh, and Guestrin}{Ribeiro
  et~al\mbox{.}}{2016}]%
        {LIME}
\bibfield{author}{\bibinfo{person}{Marco~T{\'{u}}lio Ribeiro},
  \bibinfo{person}{Sameer Singh}, {and} \bibinfo{person}{Carlos Guestrin}.}
  \bibinfo{year}{2016}\natexlab{}.
\newblock \showarticletitle{"Why Should {I} Trust You?": Explaining the
  Predictions of Any Classifier}. In \bibinfo{booktitle}{\emph{Proceedings of
  the 22nd {ACM} {SIGKDD} International Conference on Knowledge Discovery and
  Data Mining, San Francisco, CA, USA, August 13-17, 2016}}.
  \bibinfo{pages}{1135--1144}.
\newblock


\bibitem[\protect\citeauthoryear{Sallab, Abdou, Perot, and Yogamani}{Sallab
  et~al\mbox{.}}{2017}]%
        {DL-autnonmous17}
\bibfield{author}{\bibinfo{person}{Ahmad~El Sallab}, \bibinfo{person}{Mohammed
  Abdou}, \bibinfo{person}{Etienne Perot}, {and} \bibinfo{person}{Senthil~Kumar
  Yogamani}.} \bibinfo{year}{2017}\natexlab{}.
\newblock \showarticletitle{Deep Reinforcement Learning framework for
  Autonomous Driving}.
\newblock \bibinfo{journal}{\emph{CoRR}}  \bibinfo{volume}{abs/1704.02532}
  (\bibinfo{year}{2017}).
\newblock


\bibitem[\protect\citeauthoryear{Severi, Meyer, Coull, and Oprea}{Severi
  et~al\mbox{.}}{2021}]%
        {EG-Poisoning}
\bibfield{author}{\bibinfo{person}{Giorgio Severi}, \bibinfo{person}{Jim
  Meyer}, \bibinfo{person}{Scott Coull}, {and} \bibinfo{person}{Alina Oprea}.}
  \bibinfo{year}{2021}\natexlab{}.
\newblock \showarticletitle{Explanation-Guided Backdoor Poisoning Attacks
  Against Malware Classifiers}. In \bibinfo{booktitle}{\emph{30th {USENIX}
  Security Symposium, {USENIX} Security 2021}}. \bibinfo{publisher}{{USENIX}
  Association}.
\newblock


\bibitem[\protect\citeauthoryear{Shapley}{Shapley}{1953}]%
        {shapley}
\bibfield{author}{\bibinfo{person}{L. Shapley}.}
  \bibinfo{year}{1953}\natexlab{}.
\newblock \bibinfo{title}{A value for n-person games}.
\newblock
\newblock


\bibitem[\protect\citeauthoryear{Shokri, Strobel, and Zick}{Shokri
  et~al\mbox{.}}{2021}]%
        {EG-mem-inference2021}
\bibfield{author}{\bibinfo{person}{Reza Shokri}, \bibinfo{person}{Martin
  Strobel}, {and} \bibinfo{person}{Yair Zick}.}
  \bibinfo{year}{2021}\natexlab{}.
\newblock \bibinfo{title}{On the Privacy Risks of Model Explanations}.
\newblock
\newblock
\showeprint[arxiv]{1907.00164}~[cs.LG]


\bibitem[\protect\citeauthoryear{Shrikumar, Greenside, and Kundaje}{Shrikumar
  et~al\mbox{.}}{2017}]%
        {DeepLIFT}
\bibfield{author}{\bibinfo{person}{Avanti Shrikumar}, \bibinfo{person}{Peyton
  Greenside}, {and} \bibinfo{person}{Anshul Kundaje}.}
  \bibinfo{year}{2017}\natexlab{}.
\newblock \showarticletitle{Learning Important Features Through Propagating
  Activation Differences}. In \bibinfo{booktitle}{\emph{Proceedings of the 34th
  International Conference on Machine Learning, {ICML} 2017, Sydney, NSW,
  Australia, 6-11 August 2017}}. \bibinfo{pages}{3145--3153}.
\newblock


\bibitem[\protect\citeauthoryear{Simonyan, Vedaldi, and Zisserman}{Simonyan
  et~al\mbox{.}}{2014}]%
        {Whitebox-exp13}
\bibfield{author}{\bibinfo{person}{Karen Simonyan}, \bibinfo{person}{Andrea
  Vedaldi}, {and} \bibinfo{person}{Andrew Zisserman}.}
  \bibinfo{year}{2014}\natexlab{}.
\newblock \showarticletitle{Deep Inside Convolutional Networks: Visualising
  Image Classification Models and Saliency Maps}. In
  \bibinfo{booktitle}{\emph{2nd International Conference on Learning
  Representations, {ICLR} Workshop Track Proceedings}}.
\newblock


\bibitem[\protect\citeauthoryear{Springenberg, Dosovitskiy, Brox, and
  Riedmiller}{Springenberg et~al\mbox{.}}{2015}]%
        {Whitebox-exp14}
\bibfield{author}{\bibinfo{person}{Jost~Tobias Springenberg},
  \bibinfo{person}{Alexey Dosovitskiy}, \bibinfo{person}{Thomas Brox}, {and}
  \bibinfo{person}{Martin~A. Riedmiller}.} \bibinfo{year}{2015}\natexlab{}.
\newblock \showarticletitle{Striving for Simplicity: The All Convolutional
  Net}. In \bibinfo{booktitle}{\emph{3rd International Conference on Learning
  Representations, {ICLR} Workshop Track Proceedings}}.
\newblock


\bibitem[\protect\citeauthoryear{Szegedy, Zaremba, Sutskever, Bruna, Erhan,
  Goodfellow, and Fergus}{Szegedy et~al\mbox{.}}{2014}]%
        {szegedy2014intriguing}
\bibfield{author}{\bibinfo{person}{Christian Szegedy},
  \bibinfo{person}{Wojciech Zaremba}, \bibinfo{person}{Ilya Sutskever},
  \bibinfo{person}{Joan Bruna}, \bibinfo{person}{Dumitru Erhan},
  \bibinfo{person}{Ian Goodfellow}, {and} \bibinfo{person}{Rob Fergus}.}
  \bibinfo{year}{2014}\natexlab{}.
\newblock \bibinfo{title}{Intriguing properties of neural networks}.
\newblock
\newblock
\showeprint[arxiv]{1312.6199}~[cs.CV]


\bibitem[\protect\citeauthoryear{Uesato, O'Donoghue, van~den Oord, and
  Kohli}{Uesato et~al\mbox{.}}{2018}]%
        {uesato2018adversarial}
\bibfield{author}{\bibinfo{person}{Jonathan Uesato}, \bibinfo{person}{Brendan
  O'Donoghue}, \bibinfo{person}{Aaron van~den Oord}, {and}
  \bibinfo{person}{Pushmeet Kohli}.} \bibinfo{year}{2018}\natexlab{}.
\newblock \bibinfo{title}{Adversarial Risk and the Dangers of Evaluating
  Against Weak Attacks}.
\newblock
\newblock
\showeprint[arxiv]{1802.05666}~[cs.LG]


\bibitem[\protect\citeauthoryear{{Warnecke}, {Arp}, {Wressnegger}, and
  {Rieck}}{{Warnecke} et~al\mbox{.}}{2020}]%
        {explainEval}
\bibfield{author}{\bibinfo{person}{A. {Warnecke}}, \bibinfo{person}{D. {Arp}},
  \bibinfo{person}{C. {Wressnegger}}, {and} \bibinfo{person}{K. {Rieck}}.}
  \bibinfo{year}{2020}\natexlab{}.
\newblock \showarticletitle{Evaluating Explanation Methods for Deep Learning in
  Security}. In \bibinfo{booktitle}{\emph{2020 IEEE European Symposium on
  Security and Privacy (EuroS P)}}. \bibinfo{pages}{158--174}.
\newblock


\end{thebibliography}
\section*{Appendix}\label{sec: appendix}

\subsection{Baseline Attacks Hyper-Parameters}
\label{hyper}
Besides the $p$ values of the employed $||.||_p$ distances and $\epsilon$ values that we specified in section \ref{sec: eval}, the parameters used for each baseline attacks are specified in Table \ref{tab:params}:

\begin{table}[b!]

\caption{Attack Hyper-Parameters.}\label{tab:params}
 \begin{center}
    \scalebox{.80}{

   \begin{tabular}{|c|c|} \toprule
    %   \hline
      \textbf{Attack} & \textbf{Hyper-Parameters} \\ \hline
         FGS &  \shortstack{$lb = 0$, $ub = 1$,\\ which indicate the lower and upper \\ bound of features. These two parameters are\\ specified only for MNIST dataset.}\\
         \hline
    
      BIM  & $rand\_minmax=0.3$. \\
     \hline
       PGD& \shortstack{$lb=0$, $ub=1$, with solver parameters:\\ ($\eta = 0.5$, $\eta\_min=2.0$, $max_{iter}=100$)}. \\
       \hline
      MIM& \shortstack{$eps\_iter=0.06$, $nb_{iter}=10$, \\ $lb=0$, $ub=1$, $decay\_factor=1.0$)}. \\
       \hline
      SPSA & \shortstack{$learning\_rate=0.01$,   \\$spsa\_samples=128$, $nb_{iter}=10$}. \\
       \hline
      HSJA& \shortstack{$initial\_num\_evals=100$, $max\_num\_evals=10000$,\\ $stepsize\_search="geometric\_progression"$,\\ $num\_iterations=64$, $gamma=1.0$, $batch\_size=128$.} \\ \hline
   \end{tabular}}
 \end{center}
\end{table} 
\end{document}